\documentclass[aps,pre,showpacs,twocolumn,showkeys,amssymb,amsmath,nobibnotes,superscriptaddress,floatfix]{revtex4-1}
\usepackage[colorlinks=true,linkcolor=red]{hyperref}
\usepackage{amsmath}
\usepackage{amsfonts}
\usepackage{amssymb}
\usepackage{color}
\usepackage{graphicx}
\usepackage[sort&compress]{natbib}
\newcommand{\autq}[1]{\ensuremath{\text{Aut}(#1)}}
\newcommand{\sautq}[1]{\ensuremath{|\text{Aut}(#1)}|}

\newcommand{\bfv}{\mathbf{v}}

\newcommand{\Ord}[1]{\mathrm{O}(#1)}

\begin{document}

%\preprint{}

\title{Extreme lattices: symmetries and decorrelations}



\author{A. Andreanov}
\affiliation{Max-Planck-Institut f\"ur Physik komplexer Systeme\\
N\"othnitzer Str. 38, D-01187 Dresden, Germany}
\author{A. Scardicchio}
\affiliation{Abdus Salam ICTP -
Strada Costiera 11, 34151, Trieste, Italy}
\affiliation{INFN, Sezione di Trieste -
via Valerio 2, 34127 Trieste, Italy}
%\affiliation{Abdus Salam ICTP -
%Strada Costiera 11, 34151, Trieste, Italy}
\author{S. Torquato}
\affiliation{Department of Physics,\\
Department of Chemistry,\\
Program in Applied and Computational Mathematics,\\
Princeton Institute of the Science and Technology of Materials,\\
Princeton University, Princeton, New Jersey 08544, USA}

%\email[]{@ictp.it}


\date{\today}

\begin{abstract}
We study statistical and structural properties of extreme lattices, which are the local minima in the density landscape of lattice sphere packings, in $d$-dimensional Euclidean space $\mathbb{R}^d$. Specifically, we ascertain the distributions of densities, kissing numbers and numbers of symmetries of the packings across a wide range of dimensions using the stochastic Voronoi algorithm. The degree to which the packings decorrelate as well as the correlations between the density maxima as the space dimension increases is also investigated. We find that the extreme lattices decorrelate with increasing dimension, the least symmetric lattices decorrelate faster. The extreme lattices in a fixed dimension of space $d$ ($d\geq 8$) are dominated by typical lattices that have similar packing properties, like packing densities and kissing numbers, while the best and the worst packers are in the long tails of the distribution of the extreme lattices.
\end{abstract}

%\keywords{}

\maketitle

\section{Introduction}

The sphere packing problem, i.e. finding the densest arrangement of spheres in Euclidean space of a given dimension, is a classic problem. Its relevance stems from its applications in mathematics (e.g. geometry and number theory)~\citep{conway1999sphere,cassels1997introduction}, physics~\citep{parisi2006amorphous,parisi2008most,parisi2010mean,torquato2010jammed}, communication theory (communication over noisy channels)~\citep{shannon1948bell,shannon1948bbell} and combinatorial optimization~\citep{belitz2013new,goodman2010handbook,nemhauser1988integer,avis2009polyhedral}. Though it is very simple to formulate, finding exact solutions in $d$-dimensional Euclidean space $\mathbb{R}^d$ has proved to be an extremely difficult task: so far the answers are only known for $d=2$ and $3$~\citep{hales2005proof}~\footnote{It took over $300$ hundred years to prove the solution for $d=3$.}; very tight upper bounds on the maximal density were found for $d=8$ and $24$~\citep{cohn2009optimality}. The packing problem is an optimization problem and our intuition fails in high dimensions. Therefore even finding dense lattices in sufficiently high dimensions becomes a difficult problem. Nonetheless, many results have been obtained for this problem across dimensions; however, they are mostly specific to certain dimensions and no universal method to discover dense and/or the densest packings has been devised to date.

The simpler version of the problem restricts the set of the packings over which the search is performed to the set of Bravais lattices, where there is one sphere per fundamental cell. We will refer to Bravais lattice simply as a lattice unless otherwise specified. This problem admits an exact solutions by brute force enumeration of a set of special \emph{perfect} lattices as was proven by Voronoi~\citep{voronoi1908quelques}: the densest sphere packing is a perfect lattice (defined in Sec.~\ref{sec:defs}). However the enumeration procedure becomes impractical (and even intractable) very fast beyond $d=8$. Recently several approaches have been proposed to discover dense lattice packings from scratch, i.e. without any prior knowledge. All of them exploit the construction due to Voronoi. The sequential linear program of Marcotte and Torquato~\citep{marcotte2013efficient} is a direct reformulation of the Voronoi theory as a convex optimization problem. The Monte-Carlo approach of Kallus~\citep{kallus2013statistical} also exploits some elements of the Voronoi theory. Both of these methods have proved to be very efficient in discovering the densest lattice packings up to $d=20$, which turned out all to be the previously known densest lattice packings, but faced problems beyond 20 dimensions.

Two of the authors of this paper have recently proposed a stochastic modification of the Voronoi algorithm~\citep{andreanov2012random}, which allows one to explore the set of perfect lattices (introduced in Sec.~\ref{sec:defs}) in much higher dimensions than the works in~\citep{schurmann2009computational,sikiric2007classification,bremner2009polyhedral}. This modified algorithm allowed one to study perfect lattices in up to $d=19$ and rediscover all of the densest known lattice packings although, like in other approaches~\citep{marcotte2013efficient,kallus2013statistical}, the algorithm becomes less efficient as $d=20$ is approached. The aim of Ref.~\onlinecite{andreanov2012random} was to explore the set of perfect lattices and reveal the statistical properties of these packings as the dimensionality grew so we generated a big number of perfect lattices in any dimension from $d=8$ to $d=19$ (from various millions in $d=10$ to hundreds of thousands in $d=19$).

Another important activity related to the search for the densest packings is the identification of the domain of validity of the conjectured \emph{decorrelation} principle~\citep{torquato2006new}. It states that unconstrained correlations (except the one- and two-point correlation functions) vanish in high dimensions. More precisely, the principle states that unconstrained correlations vanish asymptotically in high dimensions and that the $g_n$ for any $n\ge 3$ can (up to small error) be inferred entirely from a knowledge of the number density $\rho$ and $g_2$~\citep{torquato2006new}. This is a very strong statement. Among its implications there is a new lower bound on the maximum packing fraction and the suggestion that the densest packing might be disordered in high dimensions. The decorrelation principle~\citep{torquato2006new,scardicchio2008estimates} should be realized in the limit $d\to\infty$. What does this mean for finite $d$? The simplest scenario is that in every dimension, sufficiently large, the best lattices decorrelates. A slight modification of this scenario would see the existence of an infinite set of ``special" dimensions in which the best lattice is not decorrelated, but this set is of small measure (like the prime numbers among the integers). The main question is then, ``how large is large"? Namely, how large needs $d$ to be to see the decorrelation principle in action? Shall we need $d>100$ or $d>10^6$? In this work we are limited to small $d$ ($d<19$) but we will see already interesting things occurring.

%[[\sal{I do not know what this sentence is trying to convey. Needs clarification.}. \sal{Again, I have to know what all of this means, but if special dimensions in very high d means that decorrelation does not apply, I seriously doubt that! Is this interpretation is correct, I would be uncomfortable make such statements.} ANTO: SAL, I HAVE MODIFIED THE TEXT ABOVE, SEE IF YOU LIKE IT NOW]]

The possibility of the densest packings being disordered, naturally leads to question whether there is a relation between the \emph{symmetry} of a sphere packing and its \emph{decorrelation properties}? The decorrelation principle remains a conjecture and a proof for the sphere packing problem is lacking, although it was proven in some specific cases~\citep{skoge2006packing,torquato2006exactly,zachary2011high}. The cases studied are families of disordered packings defined in all dimensions, like RSA packings~\citep{feder1980random,ziff1990kinetics,viot1993exact,torquato2002random,torquato2006exactly} as well as periodic high-dimensional generalisations of diamond and kagome lattices~\citep{zachary2011high}. A question suggests itself naturally: how do the decorrelation properties combine with other properties of a sphere packing, like packing density, in a fixed dimension? In other words, are strongly decorrelated packings denser than the less decorrelated ones? 

We have studied an important subset of perfect lattices, the set of \emph{extreme} lattices. They are local maxima of the packing fraction and their packing fraction cannot be improved by any local deformation of a lattice. The structure, number and correlations of local maxima of density is an important topic in the study of disordered systems~\citep{torquato2010robust,parisi2010mean,torquato2010jammed}. Although lattices are far from being disordered -- in anybody's intuition they are actually at the other end of the spectrum -- we believe that we can learn a large amount by applying the same methods used in the context of disordered systems~\citep{stillinger1984packing}.

The present work addresses the question mentioned above: how global symmetries (properly characterized) and densities of a sphere packing combined with its decorrelation properties for the case of extreme lattices. The relevance of extreme lattices for the lattice sphere packing problem is immediate. We find that, in general, more symmetric lattices are less decorrelated. However there are exceptional dimensions where this is not true, supporting the second scenario of the decorrelation properties behavior with increasing dimension, that we mentioned above. We also find that, lattices with equal densities are close in the space of lattices, under an appropriately defined metric.

\begin{figure}
	\includegraphics[width=0.47\columnwidth]{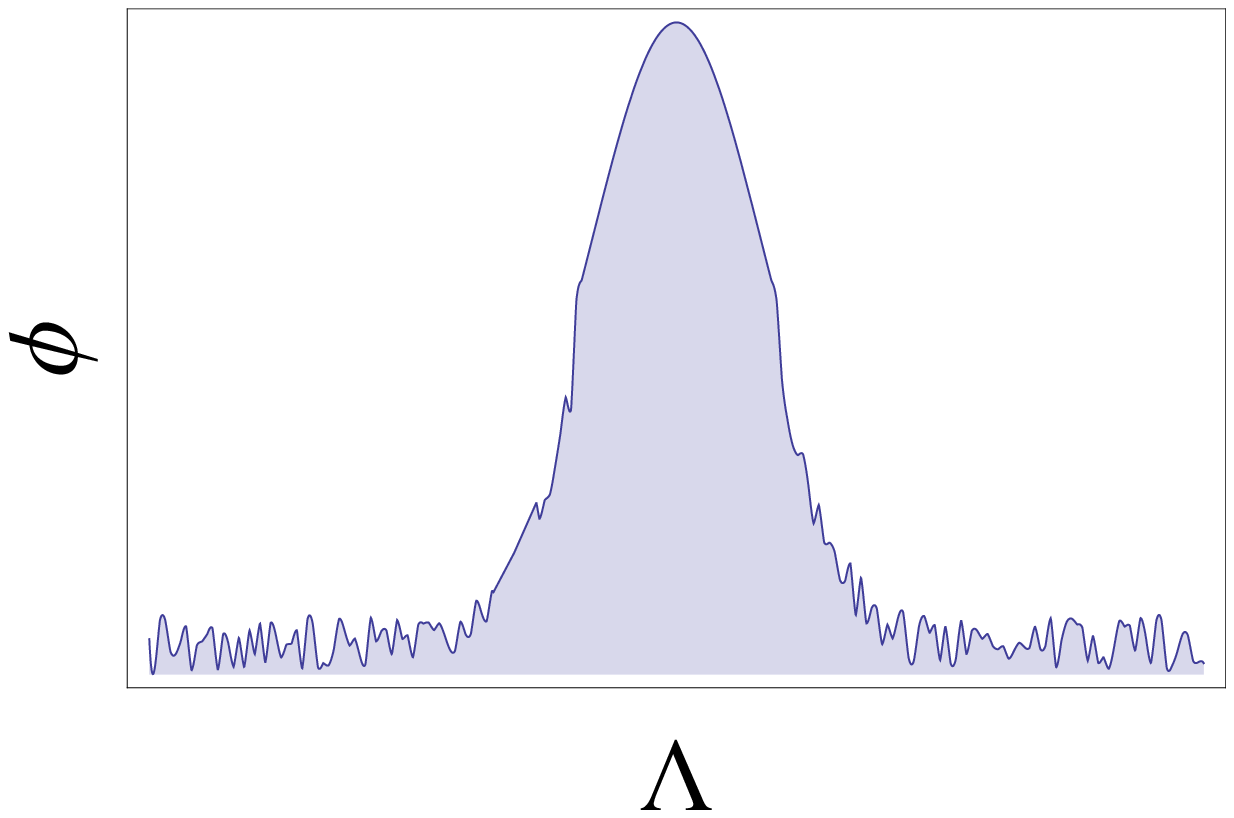}
	\includegraphics[width=0.47\columnwidth]{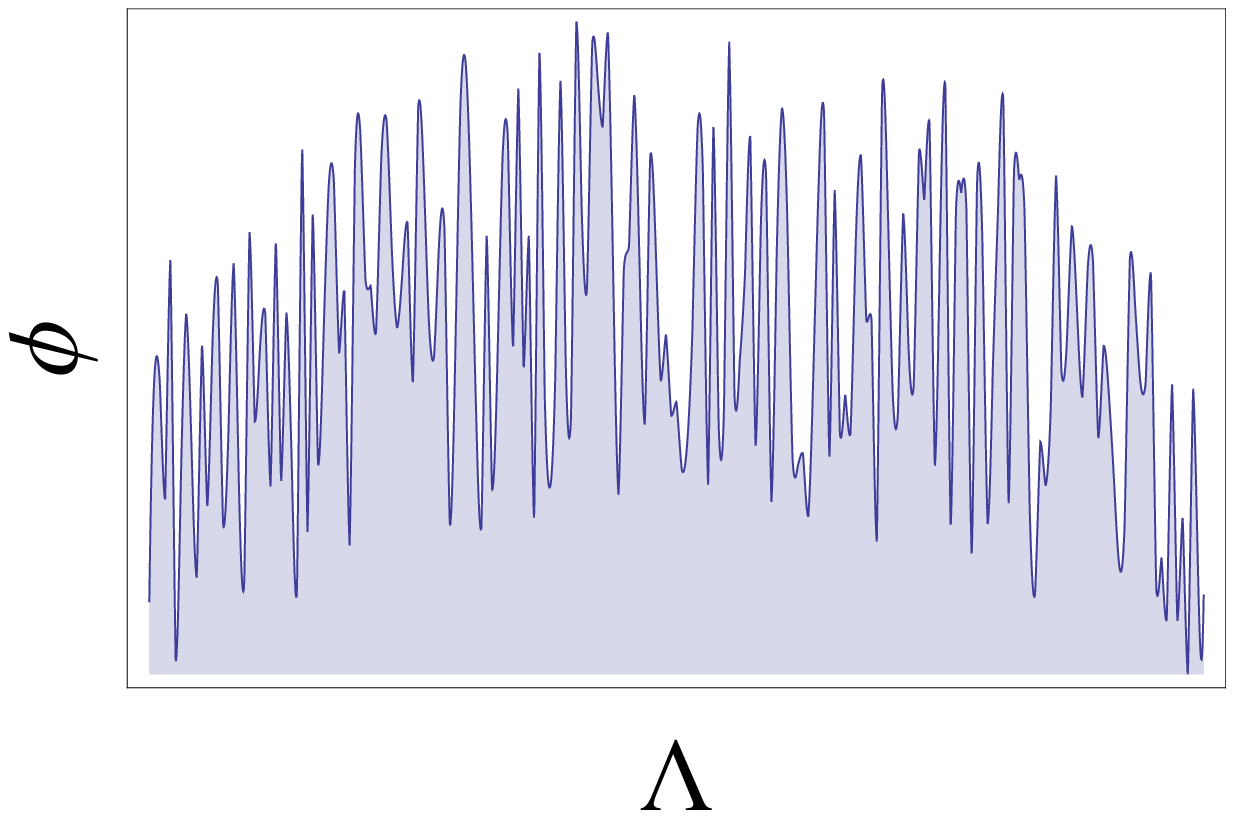}
	\caption{(Color online) Schematic representation of the density landscape $\phi(\Lambda)$ as a function of lattice $\Lambda$ for low (left) and high (right) dimensions~\citep{torquato2010robust}. In low dimensions the densest lattice has a huge basin of attraction and is easily identified. In high dimensions the basin of the densest lattices becomes comparable to that of other extreme lattices. Search for the densest lattice becomes a hard problem.}
	\label{fig:basin}
\end{figure}

We also show how the worst packing in a particular dimension $d$, which we consistently identify with the $A_d$ lattice, that is defined below\footnote{$A_d$ is a family of lattices, with one member in any dimension $d$. In 2 and 3 dimensions they are the triangular and FCC lattices and $A_d$ is the densest lattice packing up to $d=4$.}, is less decorrelated than the best packer in any dimension $d\geq 9$. This in turn, raises further questions, such as, is there a critical dimension in which best packer becomes \emph{as decorrelated as the typical} extreme lattice? Is there a critical dimension in which worst packer becomes \emph{as correlated as the typical} extreme lattice? Does the problem show features in common with configurational glasses for sufficiently high dimensions? All of the algorithms~\citep{andreanov2012random,marcotte2013efficient,kallus2013statistical} for \emph{de novo} discovery of the densest lattices suffer from noticeable performance loss around $d=20$. We attribute this to increasing complexity of the density landscape, that features many minima and the rapidly deceasing basin of attraction of the densest lattices (as schematically shown on Fig.~\ref{fig:basin}), so that the algorithms get stuck in one of the local maxima of the density -- a feature reminiscent of glassy systems. 
%[[\sal{Okay, now I see what you mean by "glassy". I have raised the possibility that the "energy" or, more precisely, "density" landscape is getting more complex, leading to a greater propensity to stuck in local density maxima. I'd like to use similar expanded language here what we mean by "glassy".} ANTO: MODIFIED]]

It is only now that we can address these fundamental questions, since we can generate a large number of extreme lattices, the magnitude of which is essentially just limited by computer time. Note that the original version~\citep{torquato2006new} of the decorrelation principle applied to \emph{disordered} systems. Recently, Zachary and Torquato~\citep{zachary2011high} extended the principle to the general class of periodic systems, including lattice packings, and we use this extension in our work. 

It is important to state at this point that the decorrelation principle is an asymptotic property of sphere packings as $d\to\infty$, i.e. very high dimensions. The dimensions accessible to us in this study, $d=8-19$ are definitely far from being high. However, the studies of decorrelation properties of sphere packings in such low dimensions~\cite{zachary2011high} indicated, that decorrelation is observed even in such low dimensions. We therefore exploit this fact, to address the questions listed above, that are otherwise very hard to study. Our hope is that at least some trends observed in such low dimensions might persist to high dimensions, or, at least, we will get some hints at what the high-$d$ behavior might be.

The layout of the paper is as follows: we introduce the main definitions that we use later in Sec.~\ref{sec:defs} and discuss the procedure to discover extreme lattices in Sec.~\ref{sec:generation}. Next we study the statistical properties of the extreme lattices, such as distributions of kissing numbers and packing fractions, and their moments, in Sec.~\ref{sec:stat_properties}. Sec.~\ref{sec:symmetries} is devoted to the study of symmetries of extreme lattices and their relation to other statistical properties of extreme lattices. In Sec.~\ref{sec:decorrelation}, we focus on the decorrelation properties of the lattices and their connection to distributions of kissing numbers and packing fractions as well as lattice symmetries. Finally, we explore the connection between the geometrical similarity and close densities for the lattices in Sec.~\ref{sec:overlap}. Our conclusions are presented in Sec.~\ref{sec:conclusions}.

\section{Definitions}
\label{sec:defs}

Here we introduce the definitions that we use throughout the paper. A lattice $\Lambda$ in $d$-dimensional Euclidean space is defined by its generator matrix $A$ or its Gram matrix $Q=A^tA$~\citep{conway1999sphere}. A lattice admits many equivalent representations in terms of the generator matrix $A$ or the Gram matrix $Q$: one can rotate the lattice or replace its basis vectors, i.e. the columns of $A$, with their independent linear combinations. This equivalence is captured by notion of \emph{isometry}: two lattices $Q$ and $Q^\prime$ are called isometric if there exists a matrix $U$ such that:
\begin{gather}
	Q^\prime = c\,U^t\,Q\,U.
\end{gather}
Another name in use is \emph{arithmetical equivalence}. For example the hexagonal lattice $Q_\text{hex}$:
\begin{gather}
	Q_\text{hex} = 
	\begin{pmatrix}
		2 & 1\\
		1 & 2\\
	\end{pmatrix}
\end{gather}
has an equivalent representation
\begin{gather}
	Q_\text{hex}^\prime = 
	\begin{pmatrix}
		2 & -1 \\
		-1 & 2\\
	\end{pmatrix}
\end{gather}
which is isometric to $Q_\text{hex}$ with isometry matrix
\begin{gather}
	U = 
	\begin{pmatrix}
		1 & -1 \\
		-1 & 0 \\
	\end{pmatrix}
\end{gather}
A practical way of checking if a given pair of forms are isometric was developed in~\cite{plesken1997computing}: one uses backtrack search to construct an isometry matrix (if this exists).

\begin{figure}
	\includegraphics[width=0.6\columnwidth]{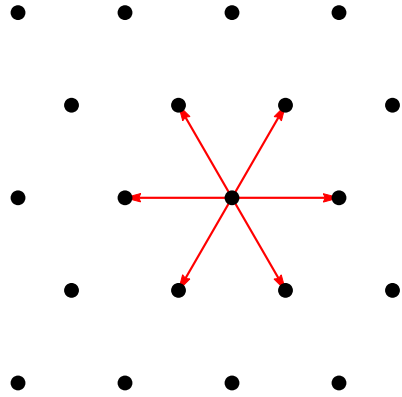}
	\caption{(Color online) The set of shortest vectors (red) of the hexagonal lattice.}
	\label{fig:hex_sv}
\end{figure}

The length of a vector represented by a set of integer numbers $\bfv\in\mathbb{Z}^d$ is given by:
\begin{gather}
	\ell(\bfv)=(\bfv,Q\bfv)^{1/2}.
\end{gather}
Lattices have vectors of all lengths, ranging from a minimum value to infinity, and an important object is the set of shortest vectors of the lattice:
\begin{gather}
	\text{Min}(Q)=\{\mathbf{v}\in\mathbb{Z}: (\mathbf{v},Q\mathbf{v}) = \min_{u\in\mathbb{Z}}(\mathbf{u},Q\mathbf{u})\}.
\end{gather}
The number of such vectors is the so-called \emph{kissing number} and is usually denoted by $Z$ as it is the number of spheres in contact with the central sphere, if we set the radius of the sphere as half the length of the shortest vectors $l(\bfv)$.  Fig.~\ref{fig:hex_sv} shows the shortest vectors of the hexagonal/triangular lattice. The scalar $\lambda(Q)=l^2(\bfv)=(\bfv,Q\bfv)$ for $\bfv\in\text{Min}(Q)$ is called \emph{arithmetical minimum} of the lattice and it is also related to the packing properties of the lattice: spheres of radius $\sqrt{\lambda}/2$ placed at the vertices of the lattice give the densest possible packing for this lattice. When talking about a lattice packing we will always imply the packing with the sphere radius $\sqrt{\lambda}/2$.

The \emph{packing fraction} $\phi$ is the ratio of the volume of the $d$-dimensional sphere of radius $\lambda/2$ to the volume of the unit cell of the lattice, given by the determinant of the basis matrix $A$:
\begin{gather}
	\phi = \frac{2^{-d}(\pi\lambda)^{d/2}}{\Gamma(1 + d/2)\det A}.
\end{gather}
Since the packing fraction decreases at least exponentially fast with the dimension~\citep{conway1999sphere}, it is convenient to define the \emph{energy}
\begin{gather}
	e = -\frac{1}{d}\ln\phi
\end{gather}
where lower energies corresponds to the better packers~\citep{andreanov2012random}.

Note that local/global maxima in the packing fraction $\phi$ are local/global minima in the energy $e$. The advantage of the energy $e$ over the packing fraction $\phi$ is its regular behavior as $d$ grows: Minkowski's lower bound on the maximal packing fraction of lattice packings scales like $2^{-d}$, while the energy remains a number of order $O(1)$. 

A lattice $\Lambda$ is called \emph{extreme} iff it is \emph{perfect} and \emph{eutactic}. \emph{Perfect} means that any symmetric $d\times d$ matrix $M$ can be expanded as:
\begin{gather}
	M = \sum_{v\in\text{Min}(Q)}\alpha_v \mathbf{v}^t\mathbf{v}.
\end{gather}
We refer the reader to the examples worked out in \citep{schurmann2009computational,andreanov2012random} to familiarize with the idea of perfect lattices.

An \emph{eutactic} lattice is one for which the identity matrix in $O(d)$ has the following decomposition:
\begin{gather}
	Q^{-1} = \sum_{v\in\text{Min}(Q)}\beta_v \bfv\bfv^T
\end{gather}
with all positive coefficients $\beta_v>0$. As was proven by Voronoi~\citep{voronoi1908quelques}, extreme lattices are local maxima of the packing fraction $\phi$. As such, they contain important information on the nature of dense packings in high dimensions.

\begin{figure}
	\includegraphics[width=\columnwidth]{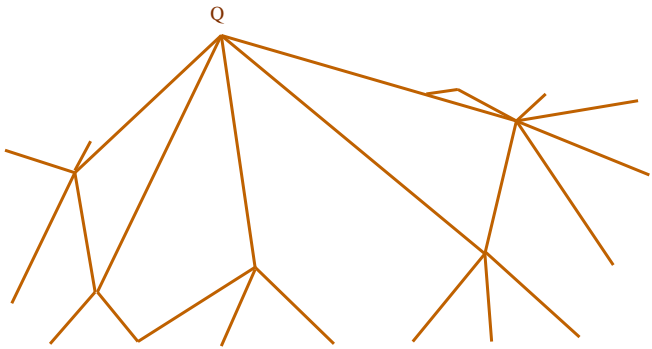}
	\caption{(Color online) A schematic drawing of a patch of the Ryshkov polyhedron defined by Eq.~\eqref{eq:ryshkov_p}. A perfect lattice defined by its Gram matrix $Q$ is a vertex of the polyhedron. The edges of the cone which has $Q$ at its top is known as the Voronoi domain of the lattice, connect $Q$ to other vertices, that are the neighboring perfect lattices.}
	\label{fig:ryshkov}
\end{figure}

The central object of the Voronoi theory is the \emph{Ryshkov polyhedron}~\citep{ryshkov1970polyhedron,gruber2007convex} 
%[[\sal{A physicist would like to see a physical meaning of this object.} ANTO: I AM NOT SURE WE CAN HAVE A DEFINITION MORE APPEALING TO PHYISICSTS. ITS A SET OF POSSIBLE LATTICES, I CANNOT VISUALIZE IT]] \alex{I added the figure}:
\begin{gather}
	\mathcal{P}_\lambda = \{Q\in\mathcal{S}_{>0}^d:\, \lambda(Q)\geq\lambda\}
\end{gather}
where $\mathcal{S}_{>0}^d$ is the set of Gram matrices of all the lattices. The definition can be rewritten in a more straightforward form:
\begin{gather}
	\label{eq:ryshkov_p}
	\mathcal{P}_\lambda = \{Q\in\mathcal{S}_{>0}^d:\, (\bfv,Q\bfv)\geq\lambda\quad\forall\,\bfv\in\mathbb{Z}^d\}.
\end{gather}
From this definitions it is clear, that $\mathcal{P}_\lambda$ is a domain, in the space of lattices, resulting from an intersection of infinite number of planes. One can prove that $\mathcal{P}_\lambda$ is convex and locally is a polyhedron~\citep{schurmann2009computational} as illustrated on Fig.~\ref{fig:ryshkov}. This is not trivial since an infinite number of intersecting planes could, in principle, produce an object that is very far from a polyhedron. Figure~\ref{fig:ryshkov} shows a patch of the Ryshkov polyhedron in $d=2$. An important result due to Voronoi asserts that the maxima of the packing fraction $\phi$ are attained at the vertices of $\mathcal{P}_\lambda$.

We also use another result due to Voronoi~\citep{voronoi1908quelques,schurmann2009computational}: the set of perfect lattices is finite and connected. Namely, for any perfect lattice one can always compute a special subset of perfect lattices, which are called its \emph{neighbors}~\citep{schurmann2009computational}. Repeating this procedure for every neighbor, one can, in principle, generate the complete set of perfect lattices. This turns the set of perfect lattices into a graph. We refer to it as \emph{the Voronoi graph} throughout the paper.

An example of extreme lattice is the $A_d$ family of lattices~\citep{conway1999sphere}. The Gram matrix of the lattice in $d$ dimensions read:
\begin{gather}
	Q_{A_d} =
	\begin{pmatrix}
		2 & -1 & 0 & 0 & \cdots \\
		-1 & 2 & -1 & 0 & \cdots \\
		0 & -1 & 2 & -1 & \cdots \\
		\cdots & \cdots & \cdots & \cdots & \cdots \\
		\cdots & 0 & 0 & -1 & 2 \\
	\end{pmatrix}
\end{gather}
Geometrically the $A_d$ lattice is defined a set of all points $\mathbf{x}\in\mathbb{Z}^d$ with integer coordinates, such that $\sum_i x_i = 0$. The $A_2$ and $A_3$ are the hexagonal, or triangular, and FCC lattices in $d=2$ and $d=3$ respectively. These are the densest lattices in respective dimensions. However beyond $d=4$ the $A_d$ are no longer the densest lattices. As space dimension increases, they become \emph{sparse}, i.e. have low density, as we will see below.

\section{Generation of sets of extreme lattices}
\label{sec:generation}

Unlike the case of perfect or eutactic lattices where algorithms exist which do, in principle, enumerate all such lattices~\citep{ash1977eutactic,voronoi1908quelques}, no algorithm is known that  generates directly extreme lattices in a sequential way (like Voronoi algorithm does for perfect lattices). One has either to start from perfect lattices and then check for eutaxy or do the opposite (the former procedure is algorithmically faster, and it is the one that we use in this paper). It is worth mentioning that two recent algorithms managed to generate random dense and maximally jammed packings~\citep{torquato2010robust,marcotte2013efficient} therefore achieving extremity without requiring separately perfectness and eutacticity. The term "random" refers to the fact that both algorithms start from a random lattice and transform it into an extreme one. Therefore they also sample the set of extreme lattices and generate its random representative. 
%\sal{These papers report truly random jammed packings, but also lattice packings from random initial conditions. Are you referring to both type of "random"?} [[ANTO: ALEX NEEDS TO ANSWER THIS]] \alex{Ok, I added tw sentences to claerify that.} 
As we said, the algorithm for generating perfect lattices and check for their eutacticity is simpler, and we have implemented it in a randomized variant in Ref.~\citep{andreanov2012random}. The number of perfect lattices is conjectured~\citep{andreanov2012random} to grow superexponentially with dimension and the number of eutactic (not necessarily perfect) lattices grows even faster~\citep{ash1977eutactic,batut2001classification}. We conjecture that the number of extreme lattices is growing at least exponentially fast with the dimension of space. Yet the fraction of perfect lattices that are also eutactic discovered by our stochastic algorithm decreases rapidly with space dimension as illustrated in Table~\ref{tab:extreme-fraction}. The increase of the fraction of lattices in $d=10-11$ is related to the bias of the algorithm towards the denser lattices and to the fact that the Voronoi graph in these dimensions is relatively small, which makes the random walk biased towards extreme lattices. In higher dimensions the size of the graph quickly negates the bias. 

\begin{table}[htbp]
	\begin{tabular}{ | c | c | c | c | c | c | c | c | c | c | c | c | c | c | }
		\hline
		Dimension & 5 & 6 & 7 & 8 & 9 & 10 & 11 & 12 & 13 \\
		\hline
		Fraction & 1 & 0.857 & 0.909 & 0.22 & 0.14 & 0.35 & 0.32 & 0.31 & 0.17 \\
		\hline
	\end{tabular}
	\caption{Fraction of perfect lattices that are also eutactic, i.e. extreme, as a function of dimension. The numbers in $d=6,7,8$ are exact, while the fractions for $d>8$ are based on the output of the stochastic algorithm~\citep{andreanov2012random}. The increase of the fractions in $d=10-11$ is likely to be related to the bias of the algorithm and is not supported by results in $d\geq 13$.}
	\label{tab:extreme-fraction}
\end{table}

These observations make generation of representative sets of extreme lattices a challenging task. Also the number of extreme lattices  discovered in a single run of the randomized Voronoi algorithm tends to be a (strongly) fluctuating quantity. There are two difficulties encountered: first, the fraction of extreme lattices (among the perfect lattices) drops sharply above $d=12$. Second, the appearance of many isometric copies of the same lattice starts becoming a problem, much more important for extreme lattices than it is for perfect lattices. Even if we bias the random walk with Metropolis-like rules by introducing an \emph{effective} temperature, we increase the fraction of extreme lattices discovered among the perfect lattices, however, for $d\geq 13$ they turn out to be mostly isometric copies of a small set of extreme lattices.

For $d\leq13$ a successful strategy is to generate sufficiently long random walks of $10^6-10^7$ steps, with a weak Metropolis bias. The instance $d=13$ seems to be a borderline case since a random walk of $6\cdot10^6$ steps starting from $A_{13}$ yielded $~2\cdot10^4$ extreme lattices. An estimate of a similar run in $d=14$ would give only a few hundred lattices with substantial increase of the running time. For $d\geq 14$ the only possibility we are left with is to perform runs at moderate temperatures, extract extreme lattices, check them for isometry and merge the set all together and perform an isometry checks over the resulting set. Following this approach, we were able to collect from $~200$ to a few thousand extreme lattices for $d=14-19$. As we show below, we cannot guarantee that lattices in such sets are representative of the typical properties of the extreme lattices. It is immediately obvious, that since the extreme lattices in these dimensions were collected from biased random walks, which favored higher density, the thus discovered extreme lattices very likely have higher densities than the typical extreme lattices discovered by simple random walks. Therefore we focus on the range $d=8-13$ and compare the results to cases $d\geq 14$ in order to check whether the sets of extreme lattices in these higher dimensions are still representative.

%As before we focus on energy and kissing number as main observables.

\section{Packing fraction and kissing number distributions}
\label{sec:stat_properties}

We start by studying statistical properties of extreme lattices. We study the same quantities - energy $e=-\log(\phi)/d$ and kissing number $Z$ - as was done in the case of perfect lattices~\citep{andreanov2012random}. 

The first issue we would like to discuss is whether statistics of energy and kissing number of extreme lattices is different from that for perfect lattices. As explained in \citep{andreanov2012random} the randomized Voronoi algorithm is hardly uniform in the choice of a neighbor of a given perfect lattice. This is due to the large variance in the dimensions of the facets of the Ryshkov polyhedron~\citep{andreanov2012random}. In order to render it more uniform we have biased the random walk. This gave a better sampling of the perfect lattices (as measured in terms of less repetitions) and the same is true for the extreme lattices as well.

\begin{figure}[htbp]
\begin{center}
	\includegraphics[width=0.9\columnwidth]{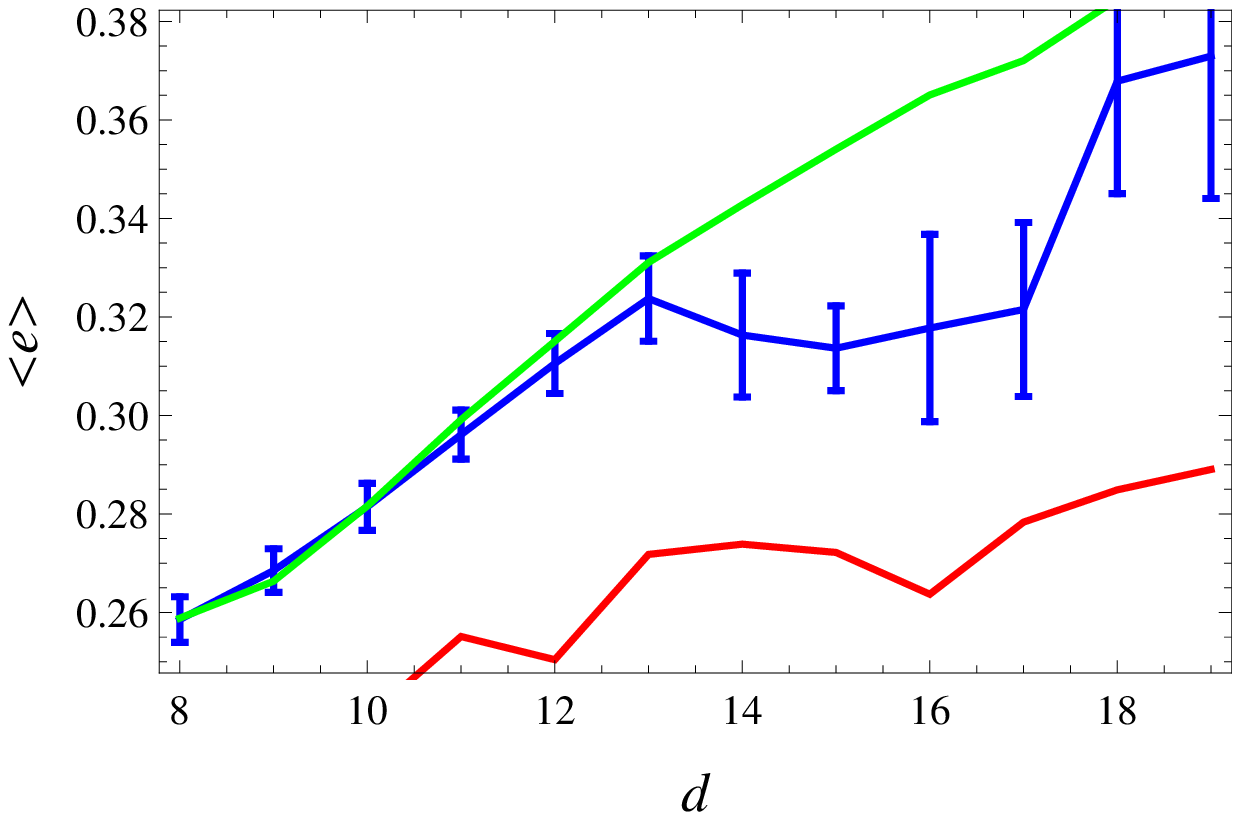}
	\includegraphics[width=0.9\columnwidth]{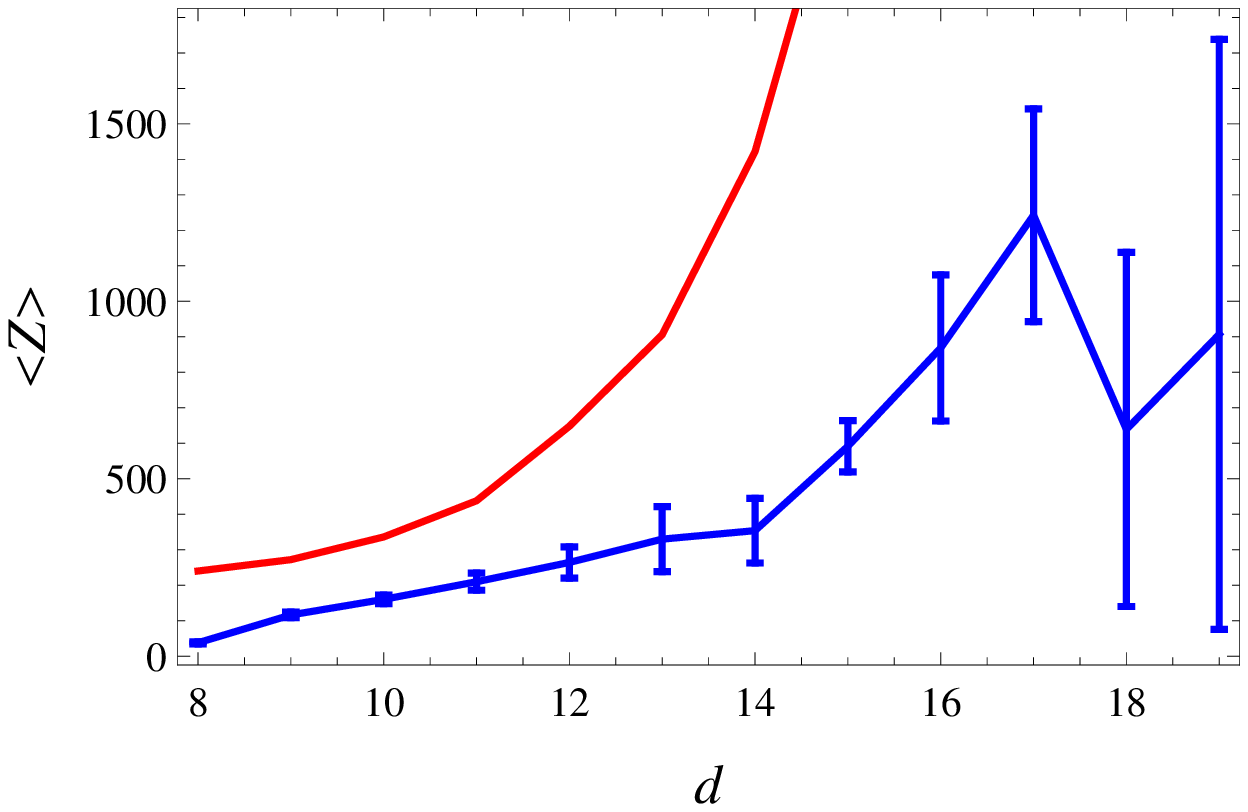}
	\caption{(Color online) \emph{Top. } Average energy of perfect (green, top curve) and extreme (blue, middle curve) lattices as a function of dimension. The red (bottom) curve is the energy of the best known packings. \emph{Bottom.} Average kissing number $\langle Z\rangle$ of extreme lattices (blue, bottom curve) as function of dimension. The red (top) curve represents the kissing numbers of the densest known packings. Error bars are variance of the distributions.}
	\label{fig:peemean8to14}
\end{center}
\end{figure}

We notice however little difference between the two algorithms and a net tendency of the extreme lattice distribution towards the densest lattices. In brief, extreme lattices are typically denser than perfect lattices (see Fig.~\ref{fig:peemean8to14}). However, it is evident that for $d\geq 13$ the sampling is far from representative and this is due both to the under sampling of the perfect lattices and to the small fraction of those which are also extreme.

The main message is that we can use the data for $d<14$ to study typical properties of extreme lattices, while in higher dimensions we can only use the results as guidelines.

At this point we notice a curios phenomenon: although in $d=3$ dimensions the lattice $A_d$ is the best packer, as the number of dimensions increases it becomes \emph{consistently} the least dense among the extreme lattices. This is a known conjecture by Coxeter~\citep{conway1988low,martinet2003perfect} and we could not disprove it in our numerics. 
%\alex{Sal: is that better ?}

\begin{figure}[htbp]
\begin{center}
	\includegraphics[width=0.9\columnwidth]{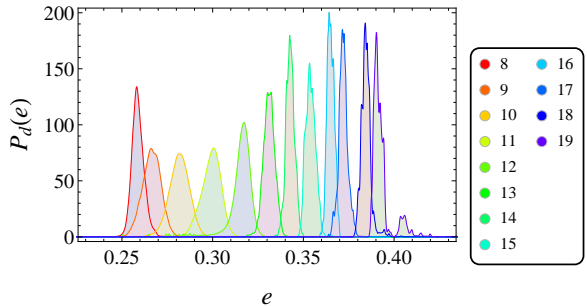}
	\caption{(Color online) Probability distributions for energy of perfect lattices in $d=8-19$.}
	\label{fig:perf-pdf8to19}
\end{center}
\end{figure}

\begin{figure}[htbp]
\begin{center}
	\includegraphics[width=0.9\columnwidth]{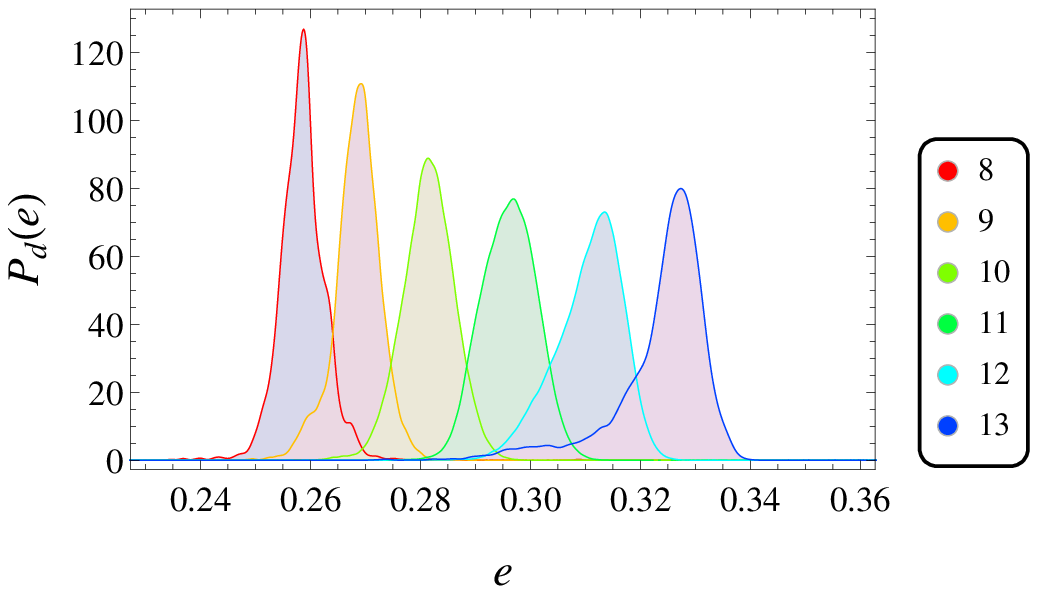}
	\includegraphics[width=0.9\columnwidth]{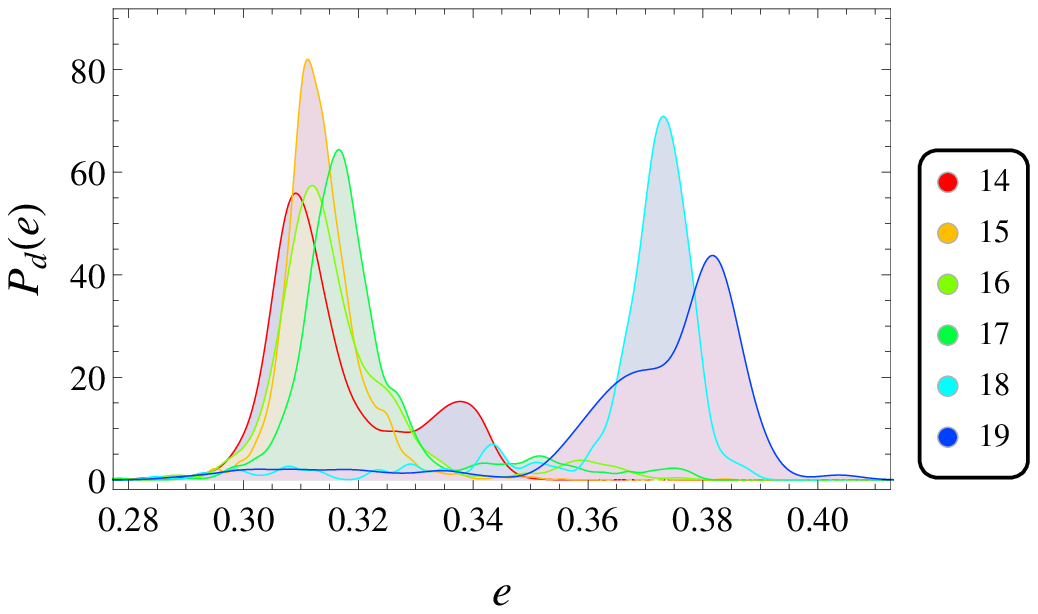}
	\caption{(Color online) Probability distributions for energy extreme lattices in $d=8-13$ (top) and $d=14-19$ (bottom). Color goes from red, $d=8$ to blue, $d=13$. Up to $d=13$ the sampling is probably good, but for $d\geq14$ we were not able to generate representative sets (see discussion in the text). The behavior for $d<14$ is similar to that of the distributions for the perfect lattices.}
	\label{fig:peepdf8to14}
\end{center}
\end{figure}

\section{Symmetries of extreme lattices}
\label{sec:symmetries}

The next problem we address, before proceeding with the decorrelation properties, are the symmetries of the extreme lattices. We need an appropriate measure of how symmetric is a sphere packing corresponding to a given lattice. The symmetries of a lattice $\Lambda$, defined by its Gram matrix $Q$, and associated packing are quantified by an \emph{automorphism group} $\autq{Q}$:
\begin{gather}
	\autq{Q} = \{U\in\text{GL}_{d}(\mathbb{Z}): U^{t}\,Q\,U = Q\}
\end{gather}
This is the set of linear, integer changes of variables in $\mathbb{Z}^n$ that map the lattice on itself, i.e. the set of all "rotations" in space, under which the lattice points map onto themselves~\footnote{If $\Lambda=A\mathbb{Z}^{d}$ and $Q=A^{t}\,A$, then every $U\in\autq{Q}$ defines an orthogonal transformation $O\in SO(d)$ such that $AU=OA$ holds.}. This is a direct probe for the symmetry of a sphere packing: higher number of symmetries implies that that packing is has higher symmetry. For example the square lattice in two dimensions has Gram matrix
\begin{gather}
	Q_{\text{sq}} =\mathbf{I}= 
	\begin{pmatrix}
		1 & 0\\
		0 & 1\\
	\end{pmatrix}
\end{gather}
and therefore the group
\begin{gather}
	\autq{Q_{\text{sq}}} = \{U\in\text{GL}_{2}(\mathbb{Z}): U^tU=\mathbf{I}\}
\end{gather}
This is the group of signed permutations, which has 8 elements in 2 dimensions (in general $2^d d!$, for $d$ dimensions).

On the other hand, the hexagonal (or triangular) lattice has Gram matrix
\begin{gather}
	Q_{\text{hex}} = 
	\begin{pmatrix}
		2 & 1\\
		1 & 2\\
	\end{pmatrix}
\end{gather}
has the following automorphism group
\begin{gather}
	\autq{Q_{\text{hex}}} = \langle
	\mathcal{A}_1=\begin{pmatrix}
		0 & 1\\
		1 & 0\\
	\end{pmatrix}
	,
	\mathcal{A}_2=\begin{pmatrix}
		0 & -1\\
		1 & 1\\
	\end{pmatrix}\
	\rangle
\end{gather}
of $12$ elements, which are generated from the above two generators of the group~\footnote{The reader can verify by himself that the group is indeed closed and has 12 elements. I helps to notice that $\mathcal{A}_{1,2}$ are roots of the identity, in particular $\mathcal{A}_1^2=1$ and $\mathcal{A}_2^6=1$.}. Therefore the hexagonal lattice has 50\% more symmetries than the square lattice (12 instead of 8) and this corresponds to our intuitive notion of the hexagonal lattice being more symmetric than the square lattice. A crude measure of this is the number of elements in $\autq{Q}$, which we denote as $\sautq{Q}$. Notice as well, that while the definition of the group relies on the Gram matrix $Q$, which is itself defined up to an isometry $V$, the $\sautq{Q}$ does not depend on $V$. Notice as well, that while the above $d=2$ case suggests that the densest lattice is also the most symmetric among the lattices, this is not so in higher dimensions, as we will see.

We adopt as a measure of the symmetry of a lattice the size of it automorphism group $\sautq{Q}$. As the size of the group can be exponentially large in $d$ for certain lattices $Q$, it is more convenient to work with its logarithm: we define the symmetry exponent $s$ of a lattice $Q$ as:
\begin{gather}
	s = \ln\sautq{Q}/d.
\end{gather}
This quantity remains of order $O(1)$ even in the case of $A_d$, which has a fairly large symmetry group in high dimensions~\footnote{As $|Aut(A_d)|=2(d+1)!$, strictly speaking $s(d)=\ln d+O(1)$ using Stirling's formula, but for such small $d$ we do not see the logarithmic growth.}. We have used the original code by Bernd Souvignier~\citep{plesken1997computing} to compute the size of the automorphism groups $\autq{Q}$.

\subsection{Symmetries}
\label{sec:symmetries:symm}

\begin{figure}
	\includegraphics[width=0.9\columnwidth]{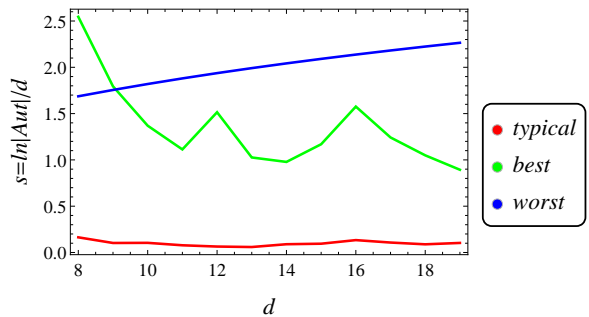}
	\caption{(Color online) The mean symmetry exponent $s=\ln|\text{Aut}|/d$ of extreme lattices (red, bottom), the best (green, middle curve) and the worst (blue, top curve) packers. The \emph{mean} curve shows smooth behavior, while the extreme cases curves are less regular. In particular the $d=12$ is special. The least dense case, which we verified being consistently $A_d$, has higher symmetry than the densest lattices for all dimensions $>8$ and asymptotically (see Appendix B) $\ln|Aut(A_d)|/d\simeq \ln d-1+\frac{3}{2}\ln d/d+...\ $.}
	\label{fig:aut-mean-extr}
\end{figure}

We first look at the distribution and the moments of $s$ for extreme lattices. Figure~\ref{fig:aut-mean-extr} shows three curves: the mean $\langle s\rangle$ over the ensemble of extreme lattices, the $s$ as a function of dimension for the best (middle curve) and the worst (top curve) packers. In all cases, the worst packer is $A_d$. We consistently find here that the best \emph{and worst} packers have high symmetry with $s\simeq 1$ for the best and $s\simeq 2$ for the worst, and there is always an exponential gap between the two. This similarity between the best and the worst can be understood as they are both the result of a global optimization (a maximum and a minimum of the same function). The trend as the number of dimensions increases is that of a \emph{decreasing} of the number of symmetries $s$ in the best packer while the number of symmetries in the \emph{worst} packer $A_d$ increases logarithmically (we have a crossing at $d=9$). Typical extreme lattices have much less symmetry, $\langle s\rangle\approx0.1$, which translates into slowly increasing $|\text{Aut}|$ from $2.23$ for $d=8$ to $6.0$ for $d=19$. This tells us that typical cases have, essentially, no symmetries, especially if compared to the best and worst cases.

\begin{figure}
	\includegraphics[width=0.9\columnwidth]{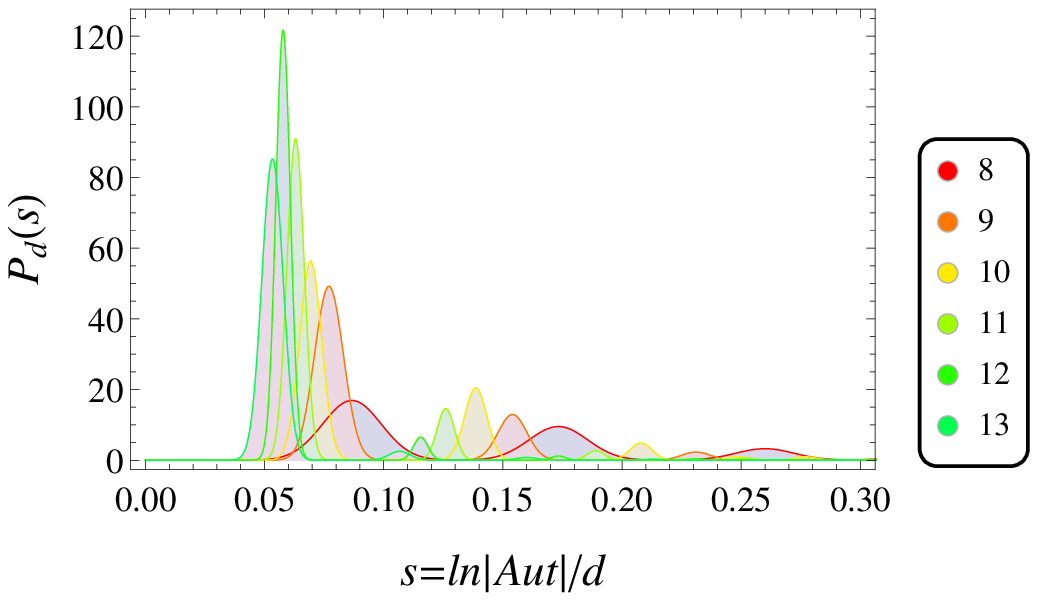}
	\includegraphics[width=0.9\columnwidth]{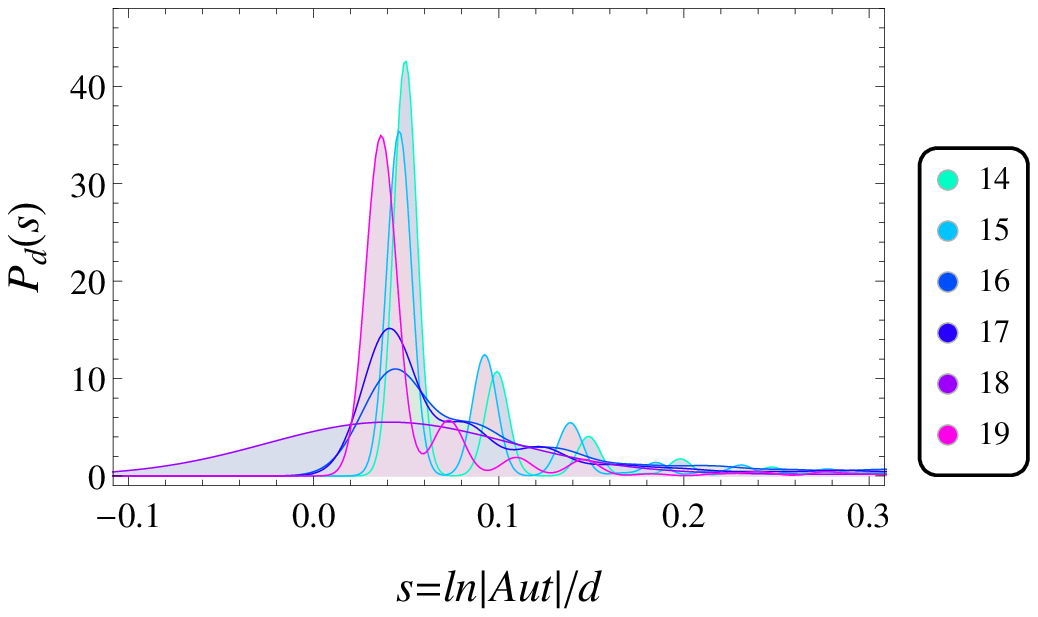}
	\caption{(Color online) The distributions of the symmetry exponent $s$ of extreme lattices for $d=8-13$ and $d=14-19$. Color gets colder as dimension $d$ is increasing.}
	\label{fig:aut-pdf}
\end{figure}

To support this statement, we study the distribution of $s$, which is shown in Fig.~\ref{fig:aut-pdf}. We show separately the cases $d=8-13$, where we have representative statistics and $d=14-19$ where we have less statistics and provide the data for illustration only. For $d<14$ the distributions of $s$ feature the main peak.

\subsection{Correlation between packing fraction, kissing number and symmetries}
\label{sec:symmetries:symm_vs_eZ}

\begin{figure}
	\includegraphics[width=0.9\columnwidth]{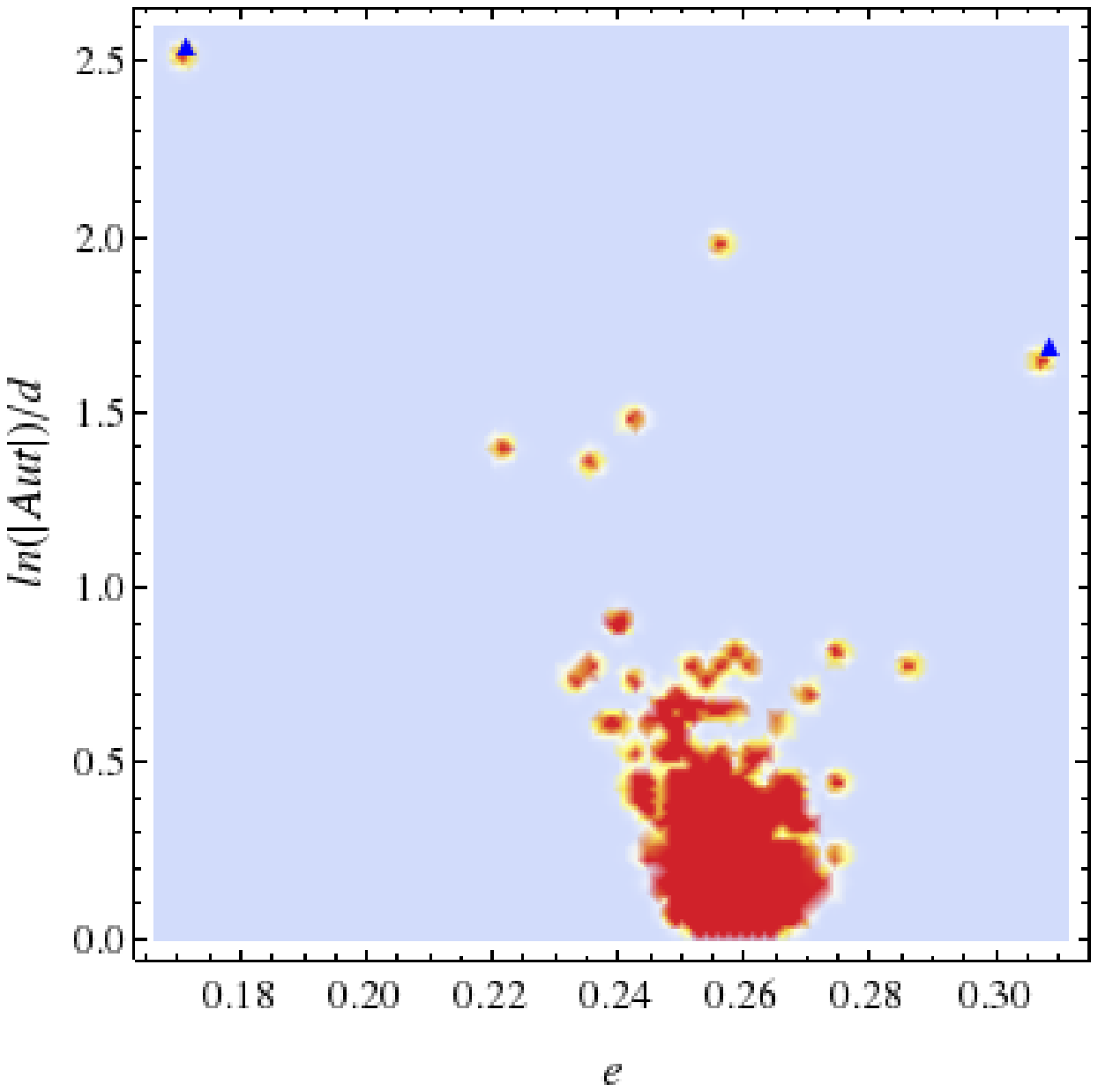}
	\includegraphics[width=0.9\columnwidth]{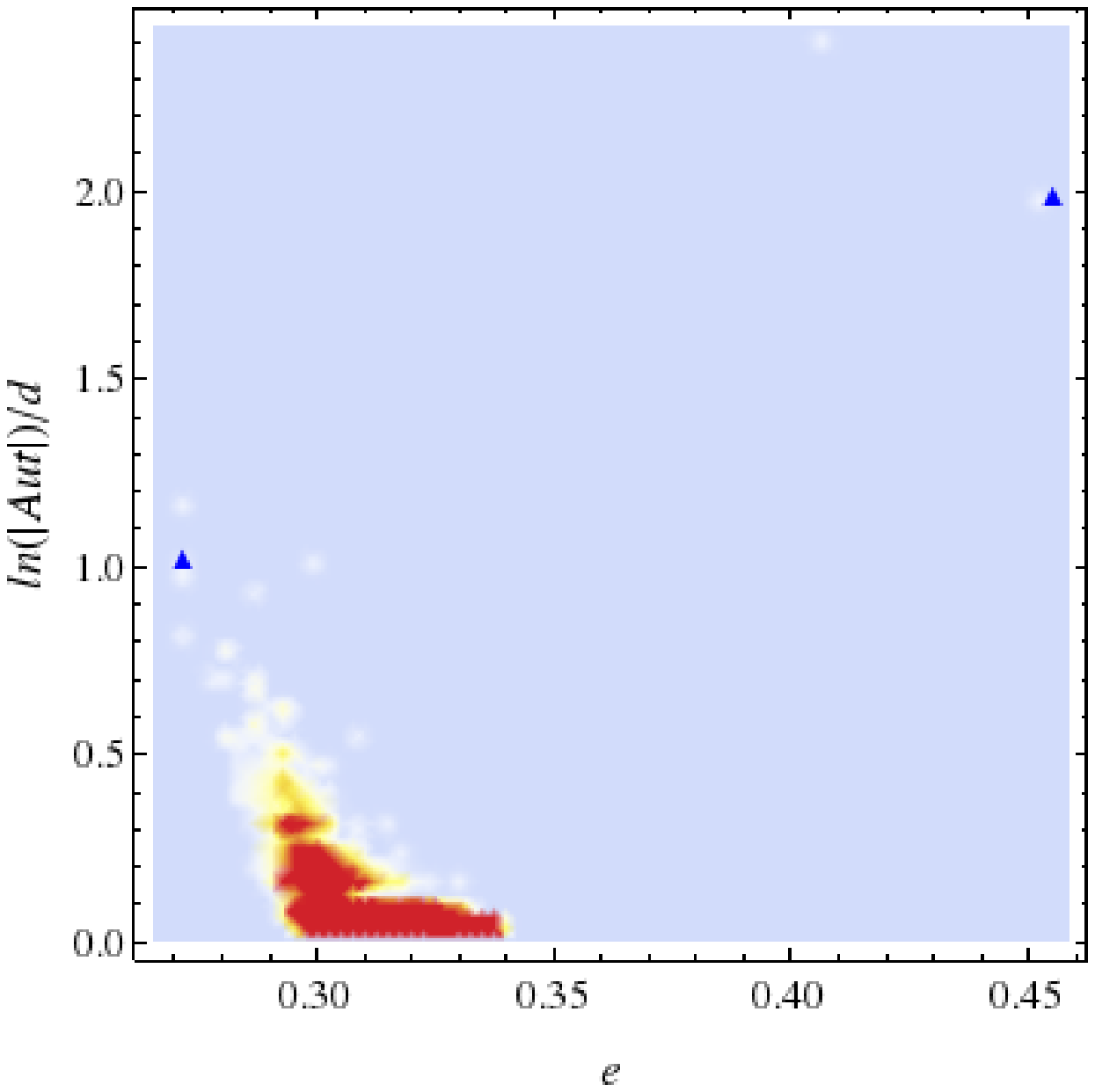}
	\caption{(Color online) The symmetry exponent $s$ of a lattice as function of energy for $d=8$ and $d=13$. The intermediate dimensions have similar scatter plots. The higher dimensions also have similar scatter plots, but their significance is reduced by the non-representative character of the sets used to generate the plots. The (red) triangular marks indicate the best and the worst packers. The densest and the sparsest ($A_d$) lattices have (much) higher number of symmetries than typical extreme lattices.}
	\label{fig:aut-vs-e}
\end{figure}

In this section we study the dependence of $\sautq{Q}$ on the lattice energy or kissing number. The results are shown in Fig.~\ref{fig:aut-vs-e}. We see that typical extreme lattices have low symmetry (small number of symmetries), sometimes as low as just $2$ transformations, while the best packers are highly symmetric. However it is not difficult to find lattices which do pretty well in packing while keeping the number of symmetries low and it is possible to find extremely high kissing numbers in packings which have $s$ value less than half that of the best kisser. This also becomes more accurate as the dimension is increased.

\begin{figure}
	\includegraphics[width=0.9\columnwidth]{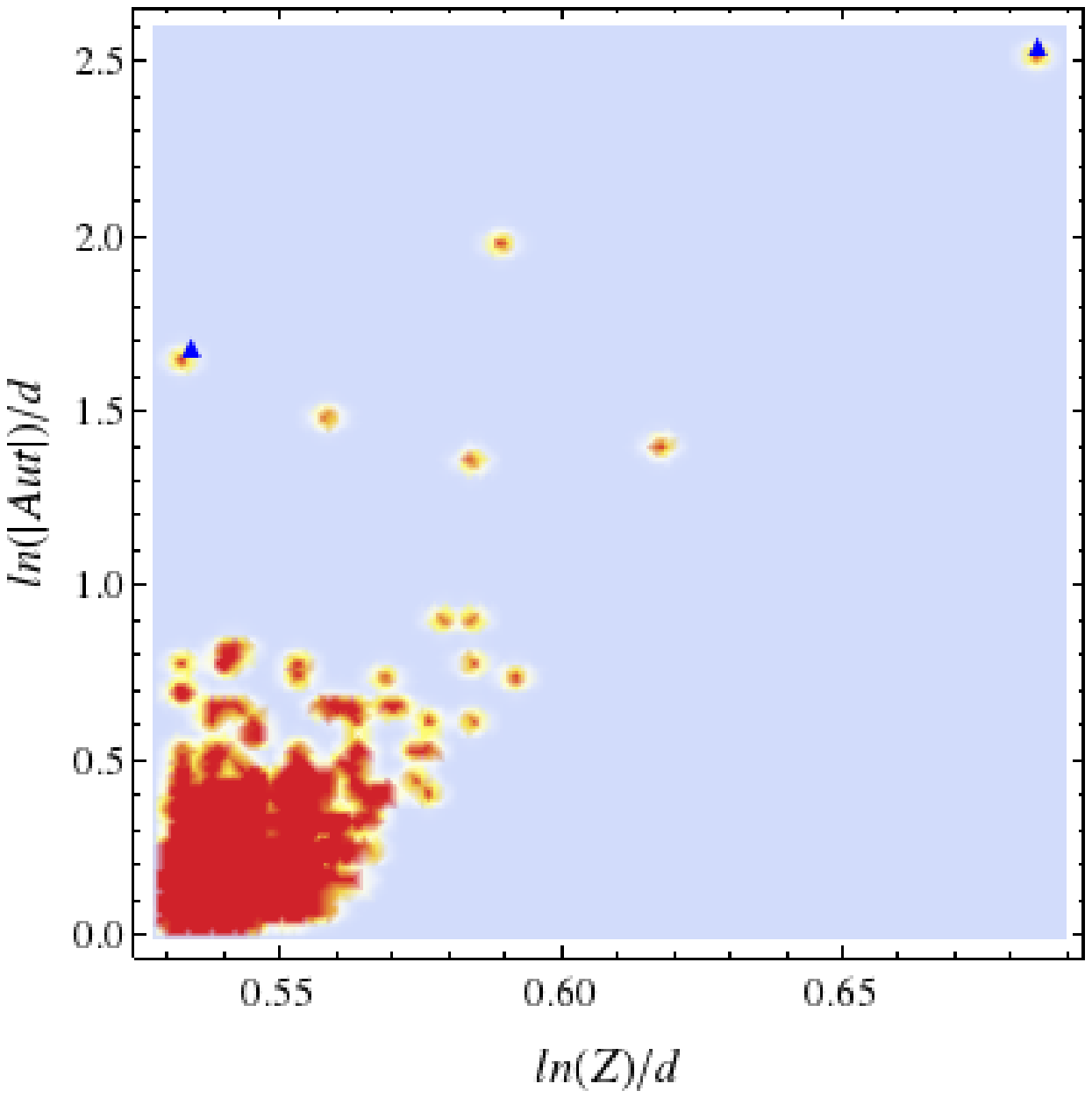}
	\includegraphics[width=0.9\columnwidth]{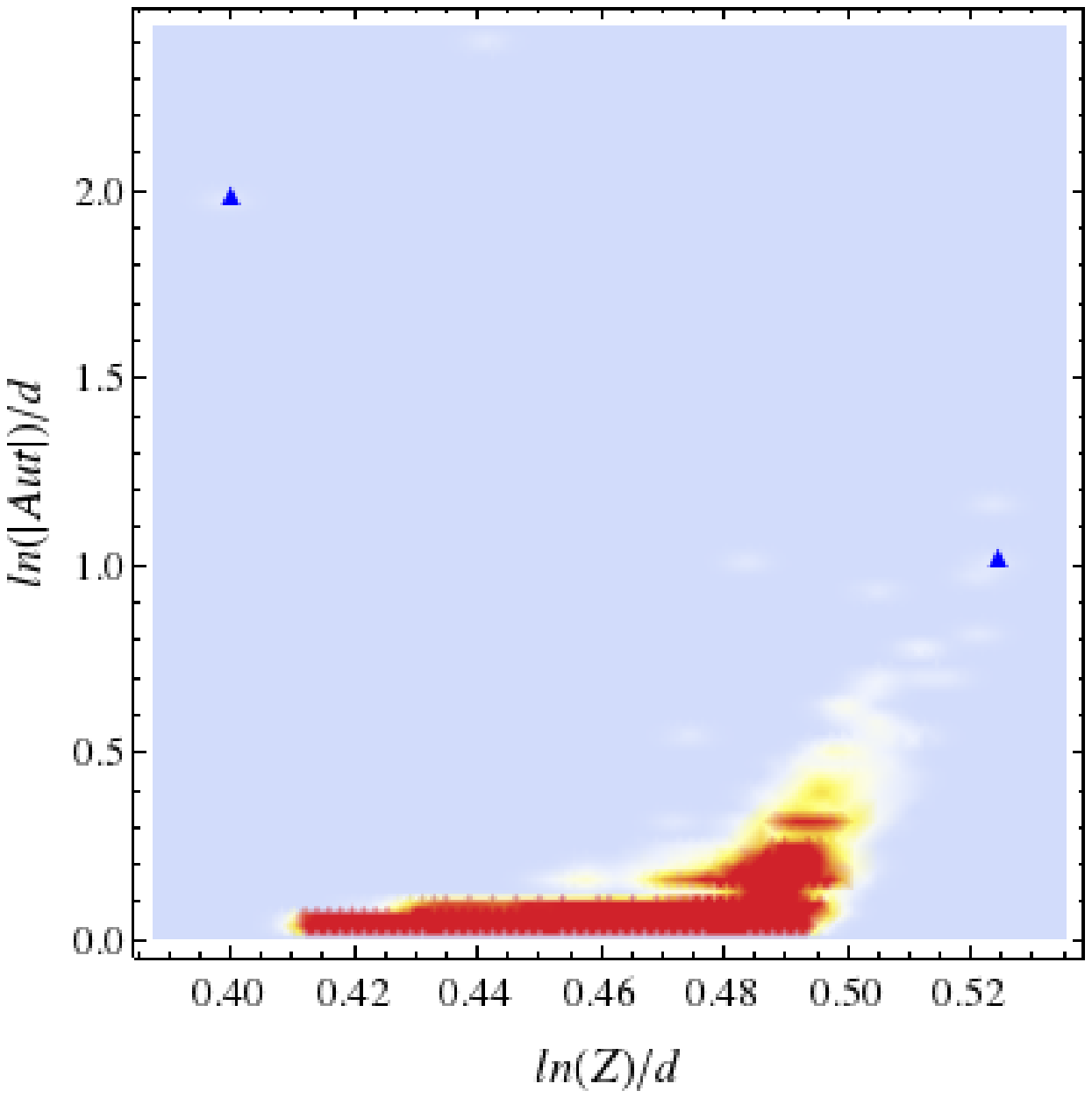}
	\caption{(Color online) The symmetry exponent $s$ of a lattice as function of the kissing number for $d=8$ and $d=13$. The (red) triangular marks indicate the best and the worst kissers. The densest and the sparsest ($A_d$) lattices have (much) higher number of symmetries than typical extreme lattices. The comments on other dimensions in the caption to the Fig.~\ref{fig:aut-vs-e} apply here as well.}
	\label{fig:aut-vs-kiss}
\end{figure}

As the dimension $d$ increases the $A_{d}$ lattices become gaped from the rest of the lattices with extremely huge symmetry groups. At the same time, they become the worst packers. A similar behavior is observed if $\sautq{Q}$ is plotted against the kissing number, as illustrated on Fig.~\ref{fig:aut-vs-kiss}.

\section{Decorrelation principle for lattices}
\label{sec:decorrelation}

We next pass on to study the decorrelation properties of extreme lattices and their connection to density and symmetry of a lattice, which is one of the main subjects of this paper. The \emph{the decorrelation principle}~\citep{torquato2006new} states that all correlations except the pair ones vanish as dimension of space is increased. If valid, the conjecture has many important consequences for the sphere packing problem \citep{torquato2006new,scardicchio2008estimates}. It has been confirmed for certain families of lattices~\citep{zachary2011high} and disordered packings~\citep{torquato2006exactly,skoge2006packing} , however its possible range of application is not well understood yet. It is therefore of interest to check if it holds for extreme lattices and study the details.

Namely, we want to study how the decorrelation properties of extreme lattices, i.e. the local maxima of the packing fraction, changes with dimension and how they depend on energy/kissing number for a fixed dimension. The latter case requires a method to compare different lattices. First, we rescale all extreme lattices to have unit length shortest vectors, i.e. we set the arithmetical minimum $\lambda=1$. We use this convention throughout the rest of the paper. This way different lattices (in a given dimension) correspond to different local arrangements of hard spheres around a central sphere. The extreme lattices, as we said, give the packings whose density cannot be improved by any infinitesimal deformation of a lattice.

%A recent advance in the comprehension of sphere packings in high dimensions is the study of the validity of the \emph{decorrelation principle}~\citep{torquato2006new}. This principle states that in high-dimensions all unconstrained correlations vanish. In particular, many-body correlations, except two-particle ones, like $g_2$, factorize. Equivalently, all the connected correlations vanish, and the only non-trivial remaining correlation is $g_2$. It is further conjectured that the structure of $g_2$ also contains less and less features as the dimension is increased. 
 
%The application of this conjecture to sphere packings gave new (conjectural) lower bound on the density of the best packings with the exponential improvement over the old Minkowski bound~\cite{torquato2006new,scardicchio2008estimates}. The other prediction following from the conjecture is the likely amorphous character of the best packers in high dimensions~\cite{torquato2006new}. Although the conjecture has been tested and confirmed in various settings~\cite{skoge2006packing,torquato2006exactly,zachary2011high}, it still remains unproven and in need of further investigations.

A direct test of the decorrelation properties of a lattice, would be to check whether higher order correlators, like the three-point correlator $g_3$, factorize into products of pair correlators $g_2$. This is computationally difficult and we set it aside for future work. A second, less direct, test of the validity of the conjecture explores the implication of the decorrelation principle, that the pair correlator $g_2$ contains less and less features and approaches $1$ for all distances as the dimension $d$ is increased. This is second possibility is feasible, instead, as the pair correlations are easier objects to compute numerically and only requires quantitative measure of features of $g_2$. More precisely, we study how the fact that $g_2$ is more or less featureless correlates with dimension/energy/kissing number as the number of dimensions is increased.

The pair correlation function $g_2(r)$ is defined as a probability of finding a particle at distance $r$ given there is a particle at the origin. The $g_2$ is also equal to derivative with respect to $R$ of the number of particles inside a sphere of radius $r$. For Bravais lattices $g_2(r)$ is set of $\delta$-functions:
\begin{gather}
	\label{eq:g2_series}
	g_2(r) = \sum_{k>0} g_k\delta(r-r_k)\\
	g_k = \frac{Z_k}{d2^d\,r_k^{d-1}\phi}\notag
\end{gather}
where $phi$ is the packing fraction, $Z_k$ is the number of lattice points in the $k$th shell and $r_k$ is the length of the lattice vectors in the shell. We refer to the series~\eqref{eq:g2_series} as \emph{$g_2$-series} in the remainder of the paper. For all lattices their $g_2(r)$ correlator can be computed exactly through the knowledge of a finite piece of their theta series~\citep{conway1999sphere,elkies2000lattices}:
\begin{gather}
	\theta_\Lambda(q) = \sum_{\bfv\in\Lambda} q^{l(\bfv)^2/2}.
\end{gather}
However the size of the required piece of the theta series can be substantial and the algorithm, based on theory of modular forms, is far from being trivial. The pair function $g_2$ has been computed by this method only for certain lattices. When this is not possible, we have resorted to a numerical method, which consists in counting all points in the lattice within a spherical region of changing radius, up to a maximum distance of few shortest vectors lengths, depending on the dimension of space. The higher space dimension, the smaller the maximum distance we have used due to computational costs.

\begin{figure}
	\includegraphics[width=0.9\columnwidth]{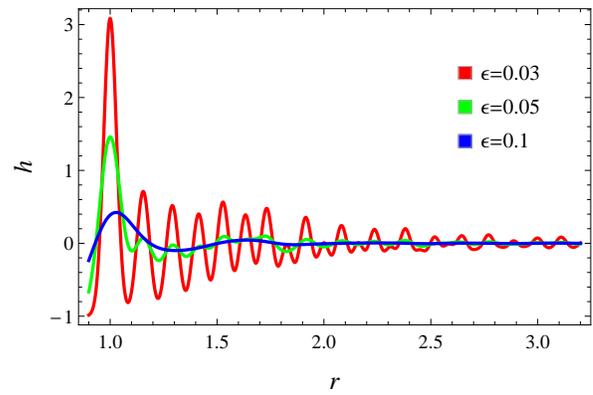}
	\caption{(Color online) The smoothed $h$ functions for a typical $d=8$ extreme lattice as defined in Eq.~\ref{eq:g2-smooth-gauss-ex} with $\epsilon=0.03,0.05,0.1$. The $\delta$-functions are smoothed into Gaussians. The correlator goes to $0$ for large distances.}
	\label{fig:g2-smooth-gauss-ex}
\end{figure}

The original decorrelation principle was developed for disordered systems: it needs modifications to be applied to lattices. The first difficulty being the obvious long-range order present which makes $g_2$ a sum of $\delta$-functions which is strictly speaking not 1 anywhere. However, recently Zachary and Torquato~\citep{zachary2011high} showed, that the decorrelation can be extended to the case of periodic systems, including  lattice packings, if one studies the \emph{smoothed} pair correlators:
\begin{gather}
	g_2(r;\epsilon) = \frac{1}{\epsilon\sqrt{2\pi}}\sum_{k > 0} g_k\exp\left[-\frac{(r - r_k)^2}{2\epsilon^2}\right]\\
h(r;\epsilon) = -1 + \frac{1}{\epsilon\sqrt{2\pi}}\sum_{k > 0} g_k\exp\left[-\frac{(r - r_k)^2}{2\epsilon^2}\right]
	\label{eq:g2-smooth-gauss-ex}
\end{gather}
where the $\delta$-functions are replaced by suitably chosen approximations (the Gaussians in the above equations), which converge to a $\delta$-function for small values of $\epsilon$. The smoothed $g_2$ should go to $1$ uniformly as $d\to\infty$, as the delta functions become denser for large $r$. An example of the smoothed $g_{2}$ correlator for a typical extreme lattice in $d=8$ is presented on Fig.~\ref{fig:g2-smooth-gauss-ex}. The effect of decorrelation is then to suppress any oscillations from $1$. In terms of $h$, we expect that
\begin{gather}
	h(r)\to 0\quad d\to\infty,
\end{gather}
for any $r>1$ (remember that we have normalized all the lattices to have unit-length shortest vectors).

Having extended the decorrelation principle to lattices, we need to define a quantitative measure of the decorrelation length of a lattice, so that comparison of different lattices is possible. To do that, we have exploited the observation that the smoothed lattice $g_2$, looses the structure and approaches $1$ on shorter and shorter distances as $d$ is increased~\citep{zachary2011high}. If one wants to define a \emph{correlation length} $\xi$ over which the smoothed $g_2$ looses its structure, one is faced with various choices. One possibility is to set $\xi = \langle r\rangle$ where the average is taken with the appropriately normalized $h^2(r)$ as a probability measure~\footnote{The choice of $h^2$ is dictated by convenience. One can use any other positive function of $h$}. Another possibility is to use the cumulative function
\begin{gather}
	\chi(r,\epsilon) = \frac{1}{\int_1^\infty dr\,h^2(r,\epsilon)}\int_1^r dr\,h^2(r,\epsilon)
\end{gather}
We set $\xi$ so that $\chi(\xi,\epsilon)=\eta$ with $\eta\sim0.9-0.99$. This last definition is close to the definition of the \emph{order metric} used to quantify disorder in materials~\citep{truskett2000towards}. Specifically, it would measure the radius (in shortest vectors) of a sphere containing $90\%-99\%$ of the order metric. 

Still another possible definition uses the cumulative function $\chi(r,\epsilon)$. One can study its behavior as $r\to\infty$ and extract a characteristic length which determines the approach to the asymptote. We have found that two different correlation lengths can be extracted with this definition from the smoothed lattice two-point correlators, describing the small-$r$ and on the large-$r$ behaviors of the $g_2$~\citep{andreanov2013astunp}.

Irrespective of the method used ($\langle r\rangle$ or $\chi$) the smaller $\xi$ the more \emph{decorrelated} is the lattice. We will see below that the decorrelation length is well correlated with symmetry (the larger the lattice symmetry group, the larger $\xi$).

It is worth stating at this point that $g_2$ does not define lattice uniquely for $d>3$. The $g_2$ series and the related $\theta$-series can be the same for different lattices, except in $d=2,3$. Counterexamples exist in $d\geq 4$, i.e. non-equivalent lattices which have identical $g_2$-series to all orders~\citep{witt1941eine,kneser1967,kitaoka1977positive,schiemann1990ein,conway1992four}. The exhaustive study of decorrelation properties of (lattice) packings would require analysis of factorization of higher-order correlators. 

\subsection{Decorrelation, energy and kissing number}
\label{sec:decorrelation:decorr_and_eZ}

\begin{figure}
	\includegraphics[width=0.8\columnwidth]{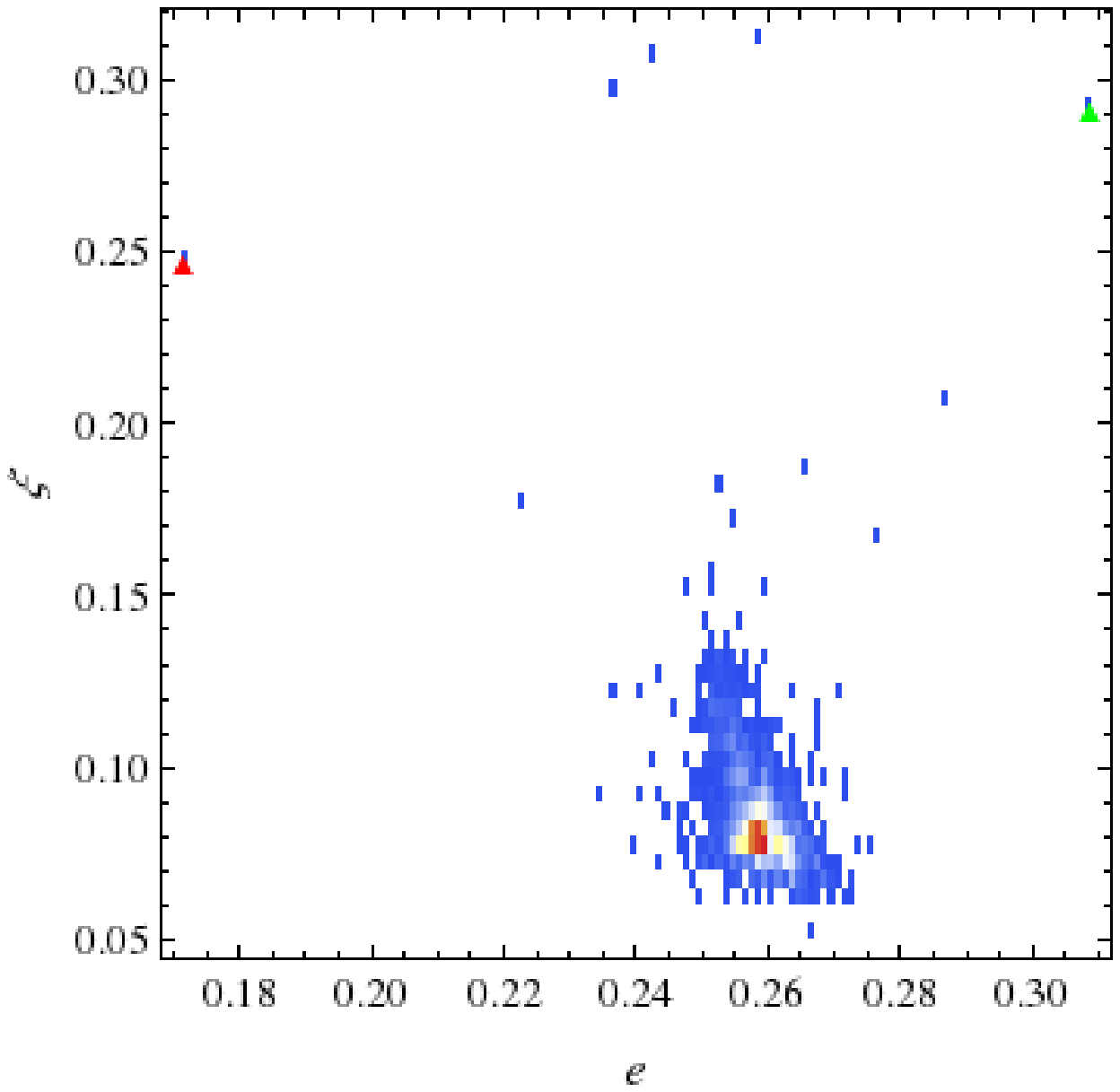}
	\includegraphics[width=0.8\columnwidth]{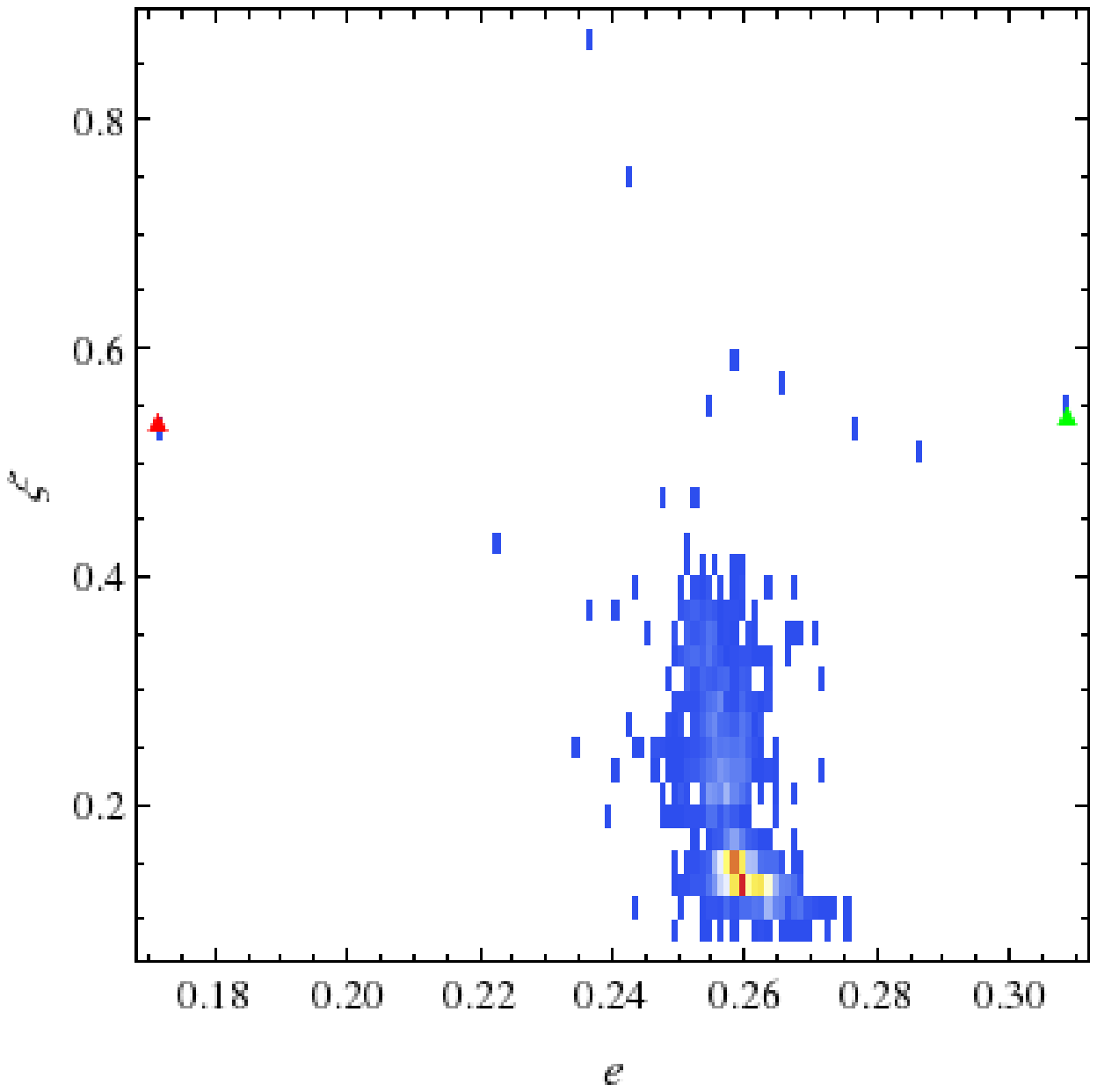}
	\includegraphics[width=0.8\columnwidth]{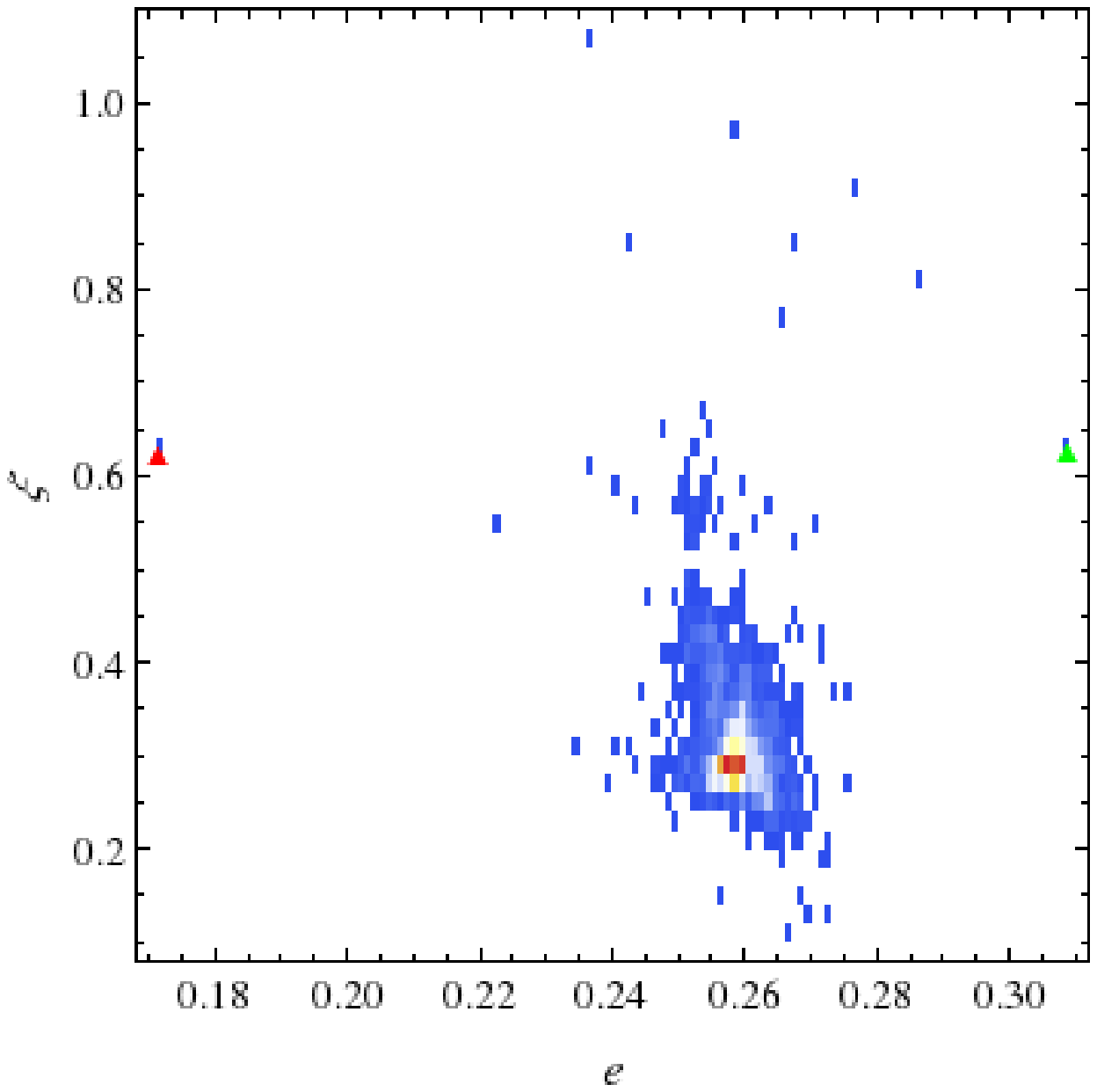}
	\caption{(Color online) Scatter plot of correlation length $\xi$ vs. energy, $d=8$. The length $\xi$ is computed with the three methods: $\langle r\rangle$ and via $h^2$ with $\eta=0.9$ and $\eta=0.95$. The results are similar qualitatively and quantitatively in this case. The relative closeness of the densest lattice in $d=13$ to the bulk of typical extreme lattices, is a peculiarity of $d=13$ as explained in the text.}
	\label{fig:rc-vs-e-d8}
\end{figure}

We first study the dependence of $\xi$ on energy and method used to compute the length $\xi$ in order to reveal the importance of the method used to extract the length. The scatter plots $\xi$ vs $e$, computed with different methods, are presented on Fig.~\ref{fig:rc-vs-e-d8}. We see that in $d=8$ different methods give very similar results. This conclusion - that the two definitions of $\xi$ are qualitatively equivalent - applies to higher dimensions as well. We use $\xi$ data obtained with the $\chi$ method with the threshold $\eta=0.95$ on the figures below.

\begin{figure}
	\includegraphics[width=0.9\columnwidth]{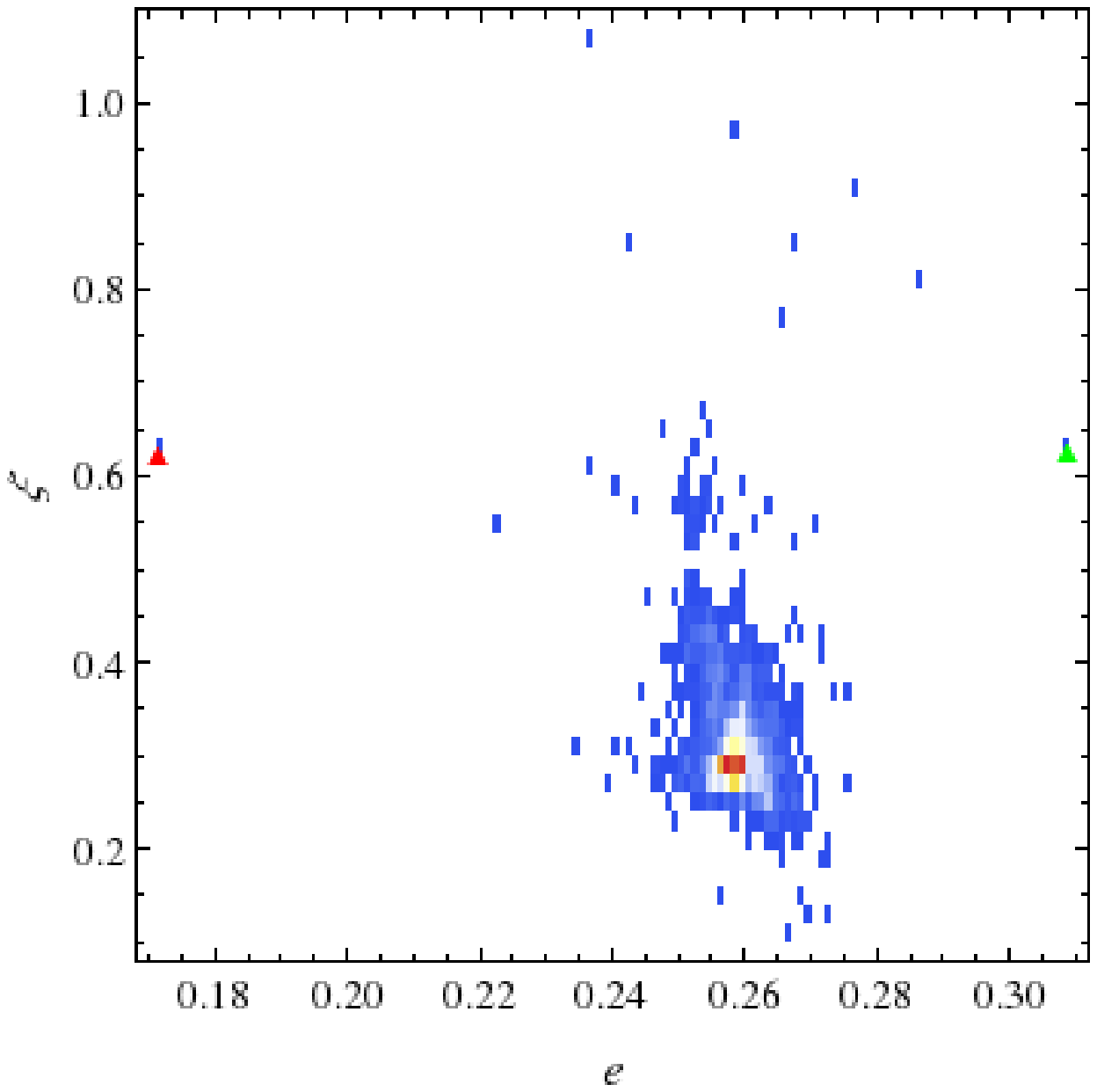}
	\includegraphics[width=0.9\columnwidth]{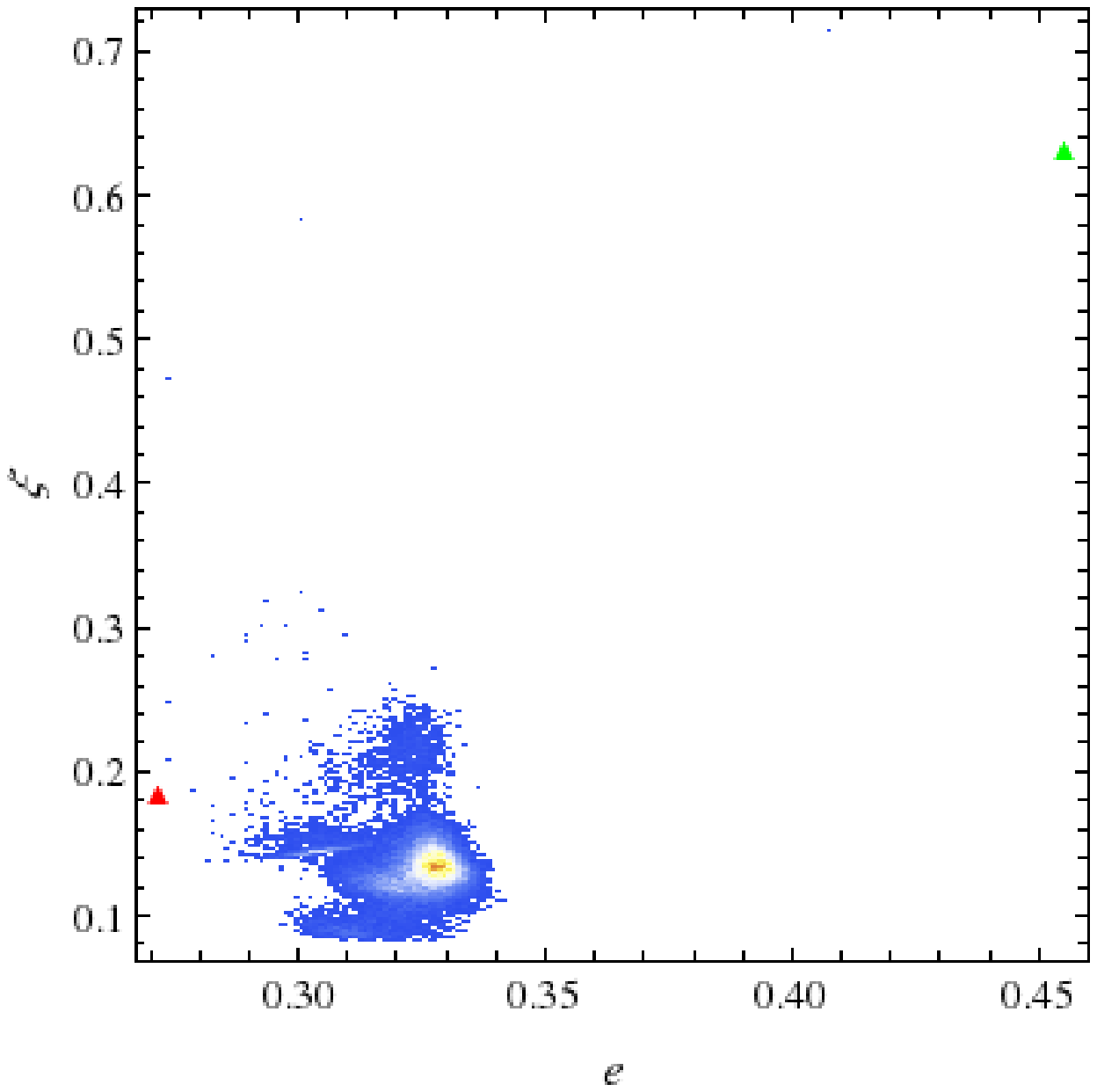}
	\caption{(Color online) Correlation length $\xi$ (computed with the  $\chi$ method, $\eta=0.95$) vs. energy $e = -\log\phi/d$ for $d=8$ and $d=13$. Typical extreme lattices have smaller correlation length in comparison with lattices having high or low energy.}
	\label{fig:rc-vs-e-h4}
\end{figure}

\begin{figure}
	\includegraphics[width=0.9\columnwidth]{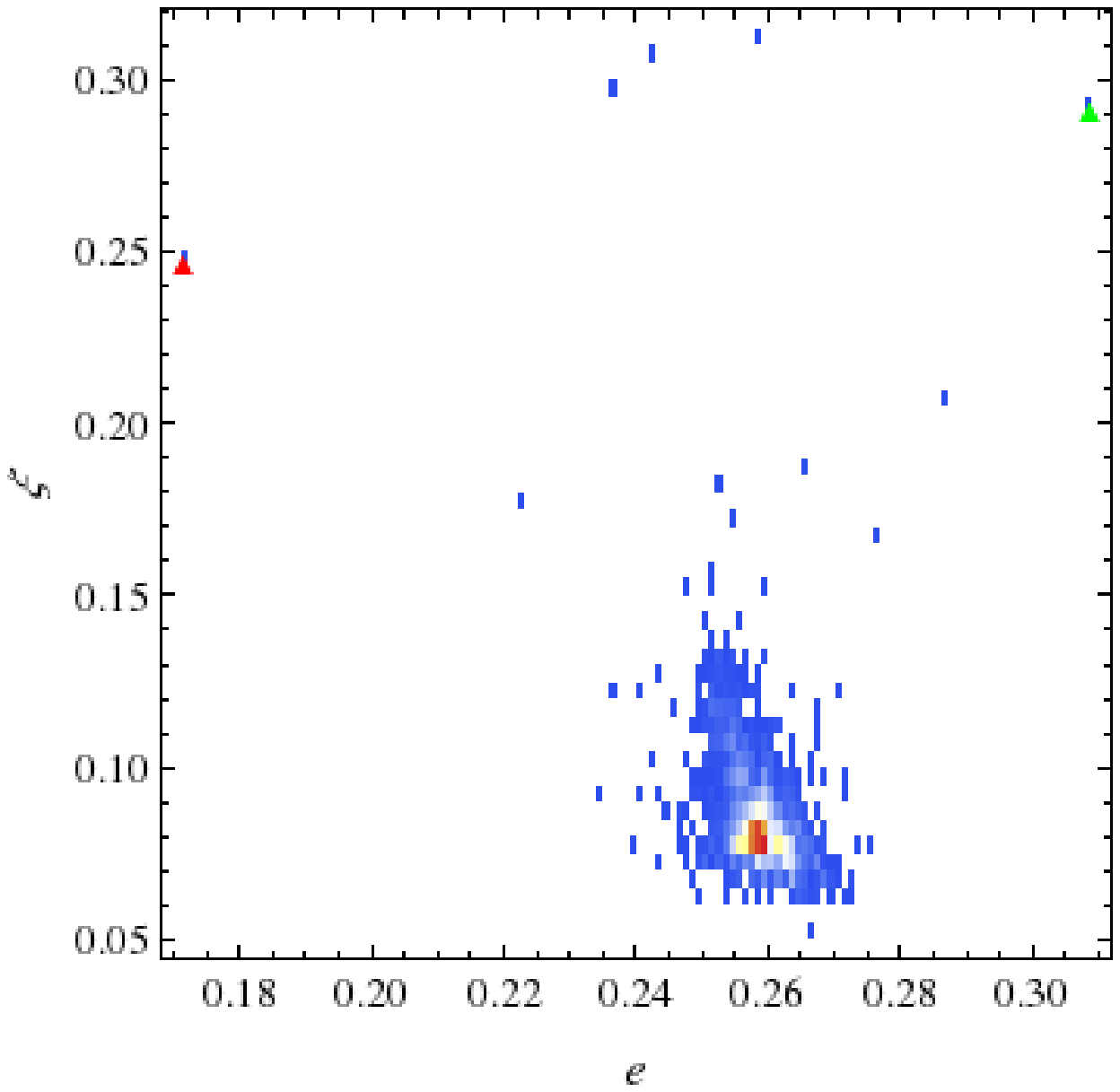}
	\includegraphics[width=0.9\columnwidth]{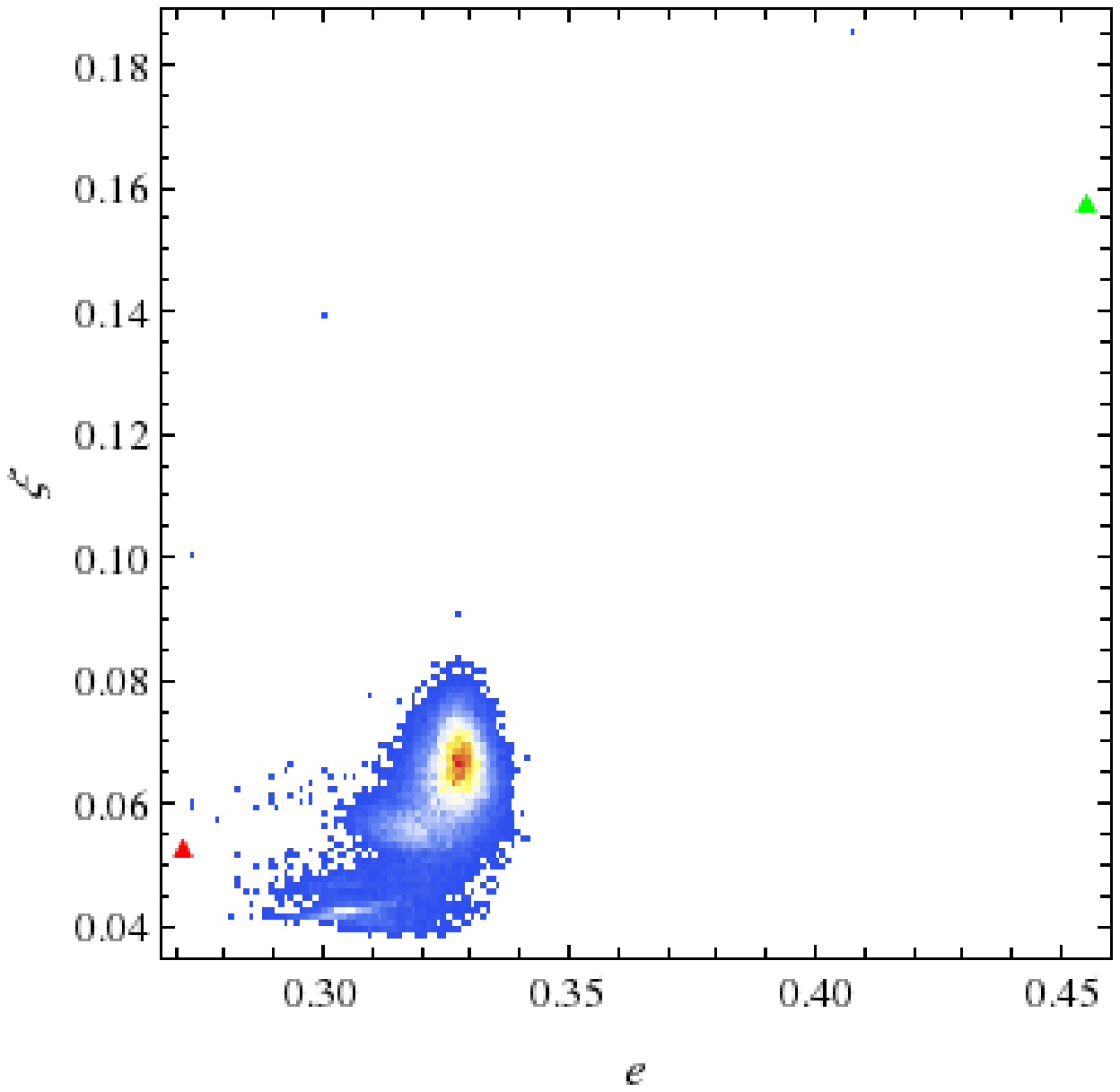}
	\caption{(Color online) The correlation length $\xi$ (computed with the $\langle r\rangle$ method) vs. energy $e = -\log\phi/d$ for $d=8$ and $d=13$. Typical extreme lattices have smaller correlation length in comparison with lattices having high or low energy for $d=8$ while the best packer in $d=13$ is less correlated than the typical extreme lattices.}
	\label{fig:rc-vs-e-r1}
\end{figure}

Looking at the scatter plots for $d=8$ and $d=13$ on Figs.~\ref{fig:rc-vs-e-h4} and~\ref{fig:rc-vs-e-r1}, we see that typical extreme lattices have smaller length $\xi$. i.e. they are more decorrelated, than the best and the worst packers. In $d=13$ the densest lattice is as much decorrelated as most extreme lattices as we see on Fig.~\ref{fig:rc-vs-e-h4}. The reason for such behavior in $d=13$ is shown in Fig.~\ref{fig:g2-smooth-best-decorr}, where the smoothed $g_2$ correlators are compared for $d=12,13,14$: the $g_2$ in $d=13$ has less structure than the smoothed $g_2$ in the neighboring dimensions.

\subsection{Decorrelation and symmetries}
\label{sec:decorrelation:decorr_and_symm}

\begin{figure}
	\includegraphics[width=0.9\columnwidth]{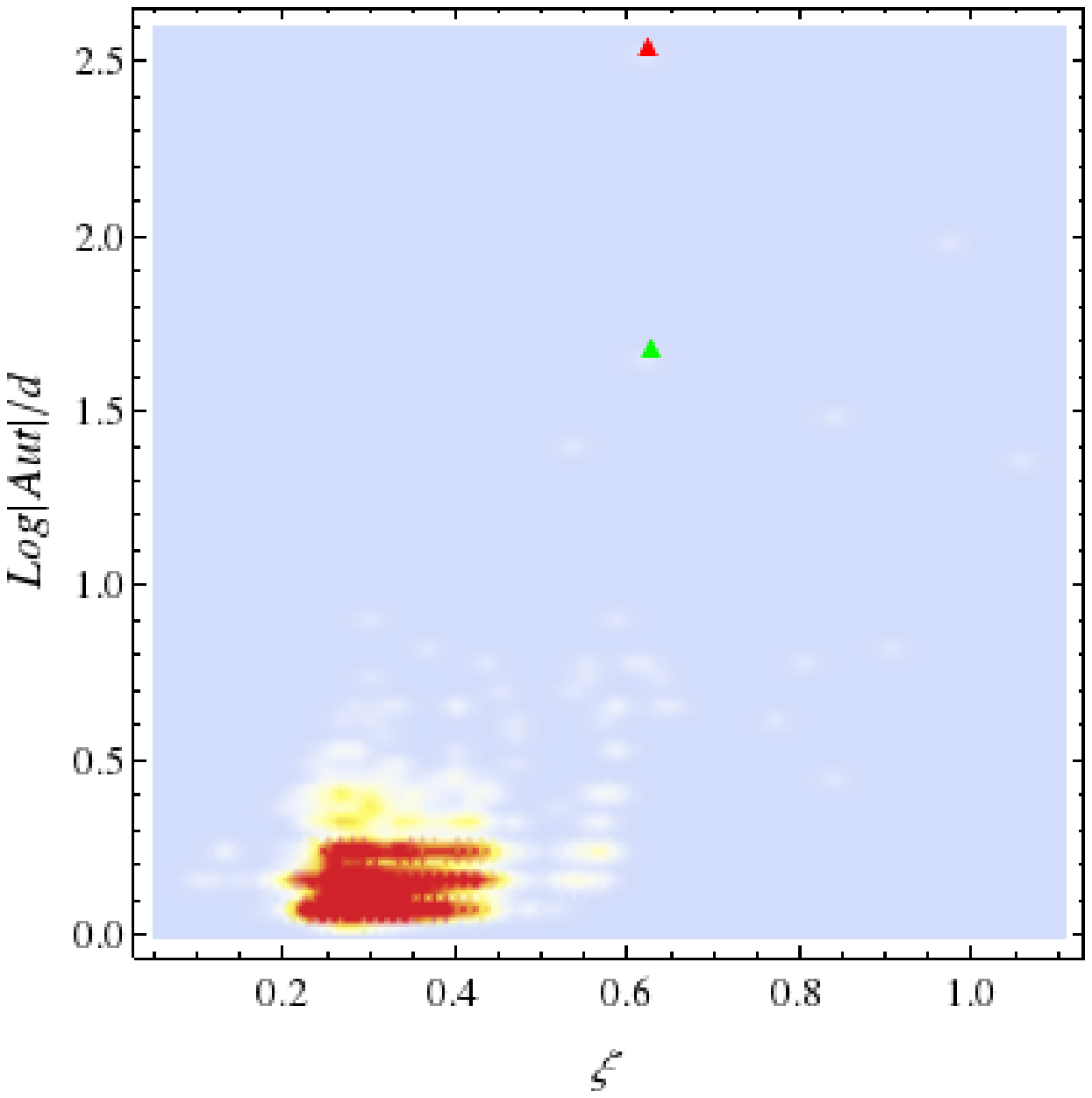}
	\includegraphics[width=0.9\columnwidth]{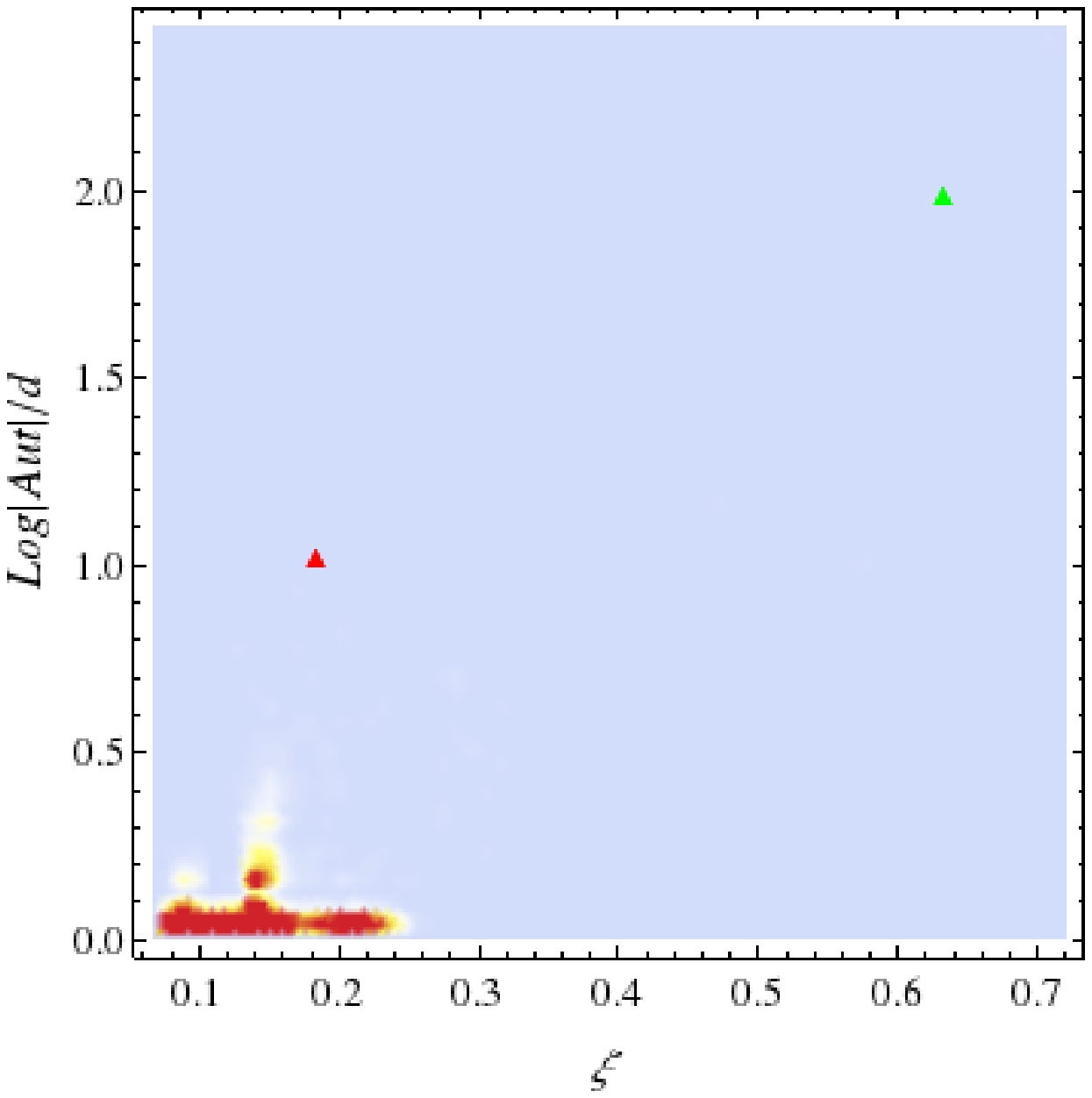}
	\caption{(Color online) Correlation length $\xi$ (computed with the $\chi$ method, $\eta=0.95$) vs. symmetry exponent $s$ of extreme lattices, for $d=8$ and $d=13$. The generic trend is that extreme lattices with lower symmetry have smaller correlation length $\xi$.}
	\label{fig:rc-vs-aut-h4}
\end{figure}

\begin{figure}
	\includegraphics[width=0.9\columnwidth]{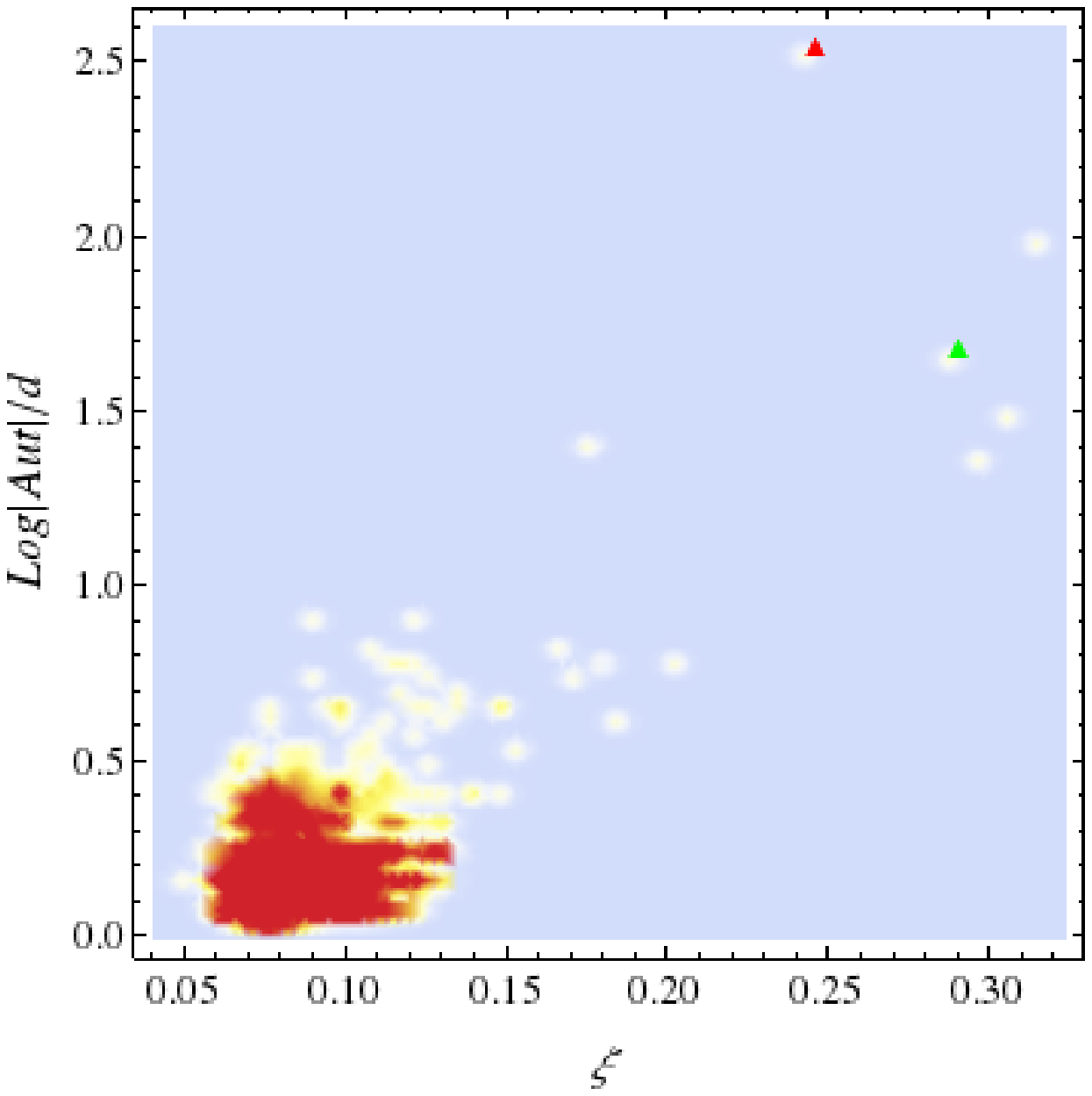}
	\includegraphics[width=0.9\columnwidth]{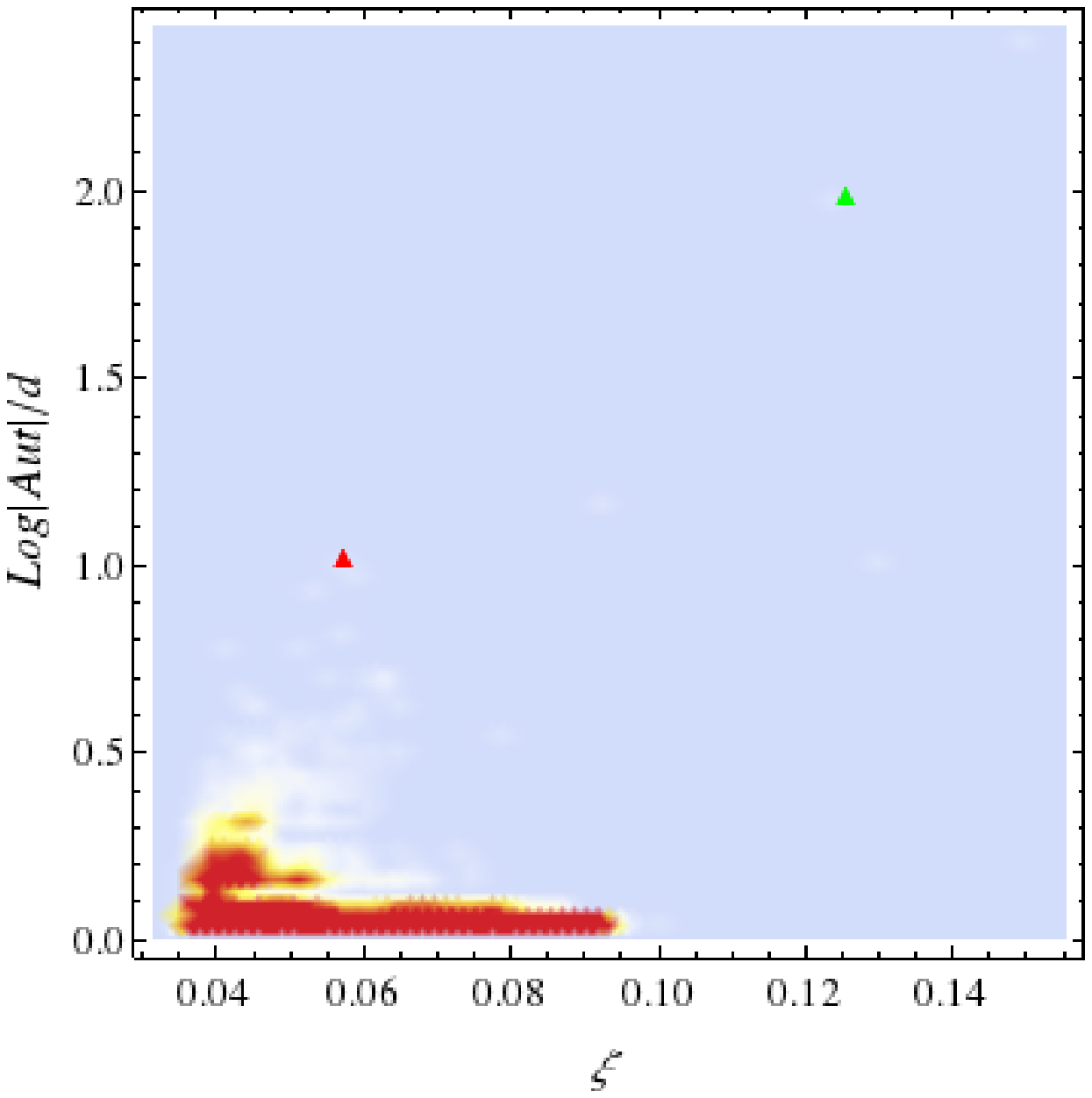}
	\caption{(Color online) The correlation length $\xi$ (computed with $\langle r\rangle$ method) vs. symmetry $s$ of extreme lattices for $d=8$ and $d=13$. Extreme lattices with lower symmetry have smaller correlation length.}
	\label{fig:rc-vs-aut-r1}
\end{figure}

Next we study the correlation between $\xi$ and the symmetry of a lattice, i.e. the size of the symmetry group of a lattice. The results are presented in Fig.~\ref{fig:rc-vs-aut-h4} (the $\chi$ method, $\eta=0.95$) and Fig.~\ref{fig:rc-vs-aut-r1} (the $\langle r\rangle$ method). We see that $\xi$ and $s$ are correlated: lattices with larger symmetry group will typically have also a larger decorrelation length. The correlation length is therefore a good measure of the (more complicatedly found) symmetry $s$.

Turning to the scatter plots themselves we see that the typical extreme lattices decorrelate faster than lattices with very high or low e number of symmetries. This fact is especially clear in case of $A_d$ or $D_d$ lattices which have huge symmetry groups already in moderately high dimensions~\footnote{$\sautq{A_{20}}=102181884343418880000$ and $\sautq{D_{20}}=2551082656125828464640000$}. This is not so in case of the best packers: the densest lattices decorrelate differently depending on the number of dimensions. In all dimensions, but $d=13$, the densest lattice is less decorrelated than typical extreme lattices. The reason the $d=13$ is special, has already been discussed in Sec.~\ref{sec:decorrelation:decorr_and_eZ}.

\begin{figure}
	\includegraphics[width=0.9\columnwidth]{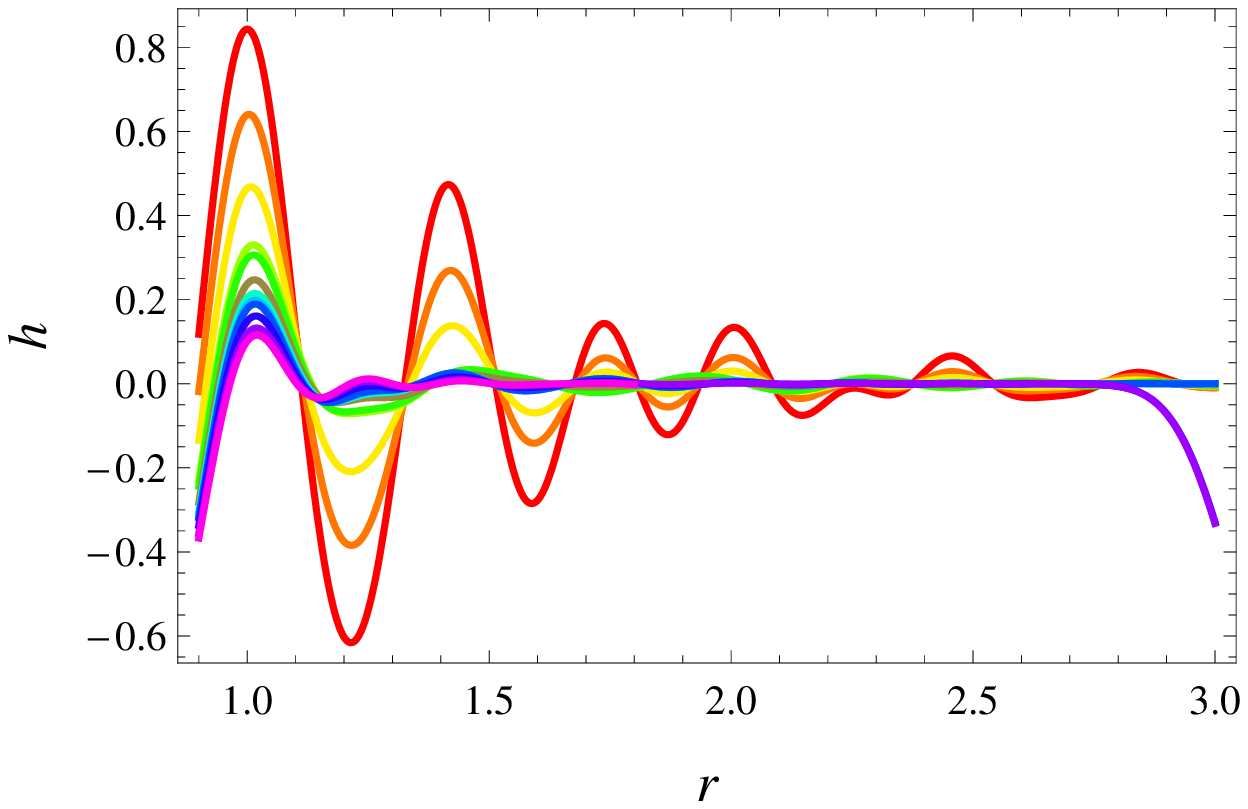}
	\includegraphics[width=0.9\columnwidth]{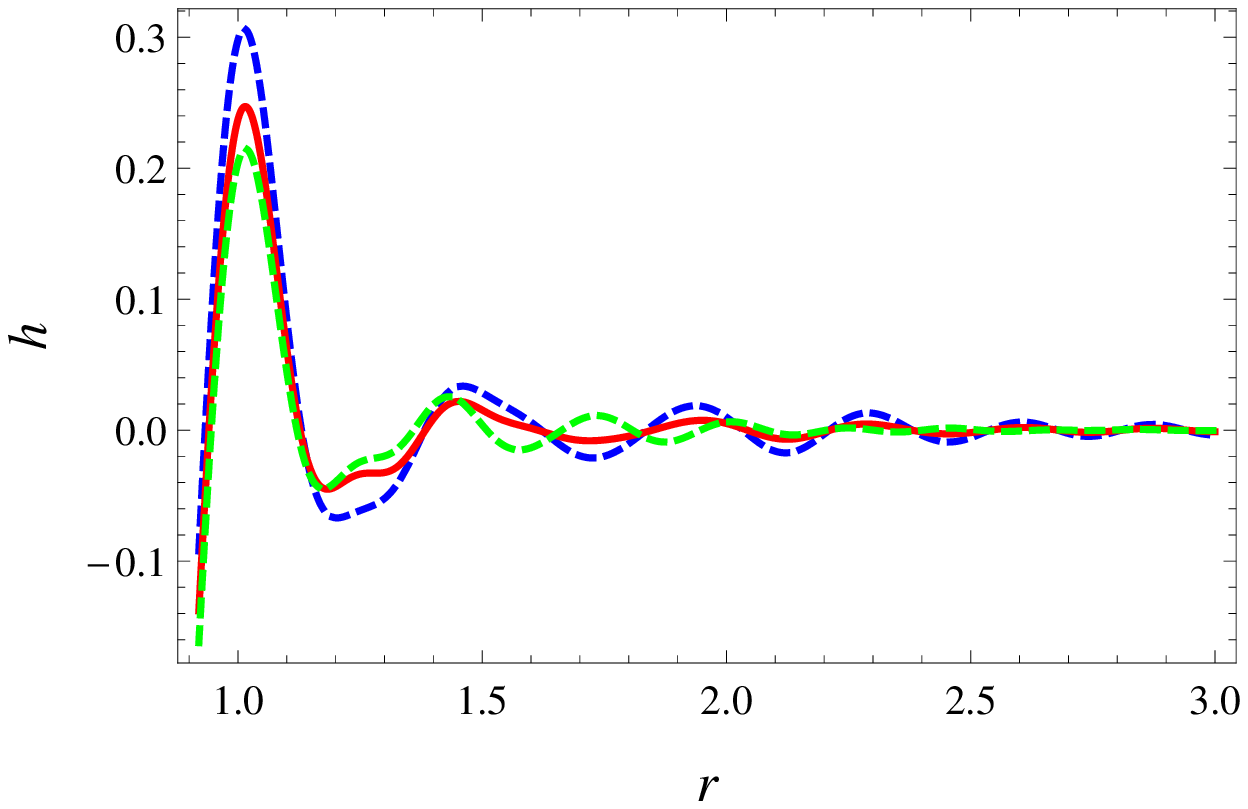}
	\caption{(Color online) \emph{Top.} The smoothed $h$ correlator (see Eq.~\eqref{eq:g2-smooth-gauss-ex}) with $\epsilon=0.1$ of the best packers in $d=8-19$. The color is darker for larger $d$. \emph{Bottom.} The same plot restricted to $d=12-14$. The smoothed $g_2$ for the best packer in $d=13$ (red, solid) has less structure than the smoothed $g_2$ in $d=12$ (blue, dashed) and $d=13$ (green, dashed).}
	\label{fig:g2-smooth-best-decorr}
\end{figure}

As the dimension increases, the best packers start to be less correlated, as one can see in Fig.~\ref{fig:g2-smooth-best-decorr} which shows the smoothed $g_2$ correlators for the best packers in $d=8-19$.
%~\alex{I wonder: in $d=12$ and $d=16$ there are particularly many  non-isometric copies of the densest lattice ? How do thet compare in terms of decorrelation ? Same question for other dimensions ?}\anto{I have no idea}.
It is evident that, although for $d=8$ the best packer is much less decorrelated than a typical extreme lattice, for $d=13$ it is already quite close to typicality and this tendency is only more evident in higher dimensions (although for the smallness of the statistics we do not present the data here).  

%\subsection{Density of states}
%\label{sec:struc_properties:dos}

%%\textcolor{red}{Does that make sense to look at binning of lattice by mutual distance ?}

%[[WHAT DO WE WANT TO DO HERE?]] \textcolor{red}{the DOS of the set of extreme lattices in terms of the lattice metric. I am not sure, though, it is worth the time.}\anto{I propose to drop this section.}

%\begin{figure}
%	\includegraphics[width=\columnwidth]{{extreme-d8.n-0-pm-0-r-8-m0.rn-100-p-2.000000-gauss-eps-0.100000.g2_dos_2}.eps}
%	\caption{(Color online) The histogram of distances of extreme lattices from $E_8$ in $d=8$}
%\end{figure}

\section{Correlations between lattices and glassiness}
\label{sec:overlap}

We have seen that typical extreme lattices in moderately high dimensions (and we conjectured that this becomes more and more true as the number of dimensions is increased) are quite homogeneous as to what concerns packing fraction and symmetries. A series of natural questions arise: what other common features are there of typical lattices? Are best packers so different from the typical ones that we could single them out by using a different metric than the packing fractions? Do they clusterize in some appropriate sense, resembling local minima of the free energy of mean-field glasses?

To answer these questions, a good starting point is to analyze whether the lattices that are close in energy are also similar in real space. The latter requires a definition of  distance $\rho(\Lambda,\Lambda')$ between lattices $\Lambda,\Lambda'$ as a measure of geometric similarity. 

Defining a workable metric in lattice space is a non-trivial problem as there might be very different presentations of the same lattice. Since lattice can be represented by many equivalent Gram matrices, the latter cannot be used tout-court to construct the metric (see Appendix for more details) and one has to scan for different presentations of the same lattice. 

An alternative possibility is to use theta series associated to a lattice~\citep{conway1999sphere}:
\begin{gather}
	\theta_\Lambda(q) = \sum_{v\in\Lambda} q^{|v|^2/2}.
\end{gather}
It is possible to define a distance in the space of lattices with the help of theta series, however its computation represents a serious mathematical problem (see Appendix B for discussion). One could think of computing the distance between $\theta$'s in function space.

The idea is good but computing the $\theta$, as we said, is cumbersome. We could implement the same idea (measuring the distance between lattices from the functional distance between their associated functions) by means of another lattice quantity, which we have already seen in this work, the smoothed pair correlation function $g_2(r,\epsilon)$. 

Based on this function we define the distance:
\begin{gather}
	\label{eq:rhopAB}
	\rho_p(A,B;\epsilon) = \left(\int\limits_1^\infty dr |g_2^A(r,\epsilon) - g_2^B(r,\epsilon)|^p\right)^{1/p}.
\end{gather}
As we have already stated, $\rho_p$ is not a metric since $g_2$ does not fix a lattice uniquely and different lattices can have identical $g_2$.

Since we believe that, as the dimension of the space increases, the knowledge of higher order correlations ($g_3$ etc) becomes less and less important (as stated by the \emph{decorrelation principle}), we propose to trust that the distance between lattices given by Eq.~\eqref{eq:rhopAB} captures the ``geometrical distance" in a reasonable sense.

There are two numerical issues: we can never compute the entire $g_2(r)$ for arbitrary distances and have to stop at some finite cutoff distance. However we have made sure that the smoothed $g_2$ is already close to its asymptotic value of $1$ at these cutoff distances. We have also checked that the smoothing does not affect the results, qualitatively.

%To illustrate these points we have computed distances between $E_8$, $A_8$, $D_8$ using theta series based distance $\tilde{\rho}_p$ and $\rho_p$ of Eq.~\eqref{eq:rhopAB} ($\epsilon=0.1$):
%\begin{gather*}
%\rho_p(E_8,A_8) = 0.088\quad\tilde{\rho}_p(E_8,A_8) = \\
%\rho_p(E_8,D_8) = 0.074\quad\tilde{\rho}_p(E_8,D_8) = \\
%\rho_p(A_8,D_8) = 0.058\quad\tilde{\rho}_p(A_8,D_8) = \\
%\end{gather*}
%These results indicate that.

\begin{figure}
	\includegraphics[width=0.9\columnwidth]{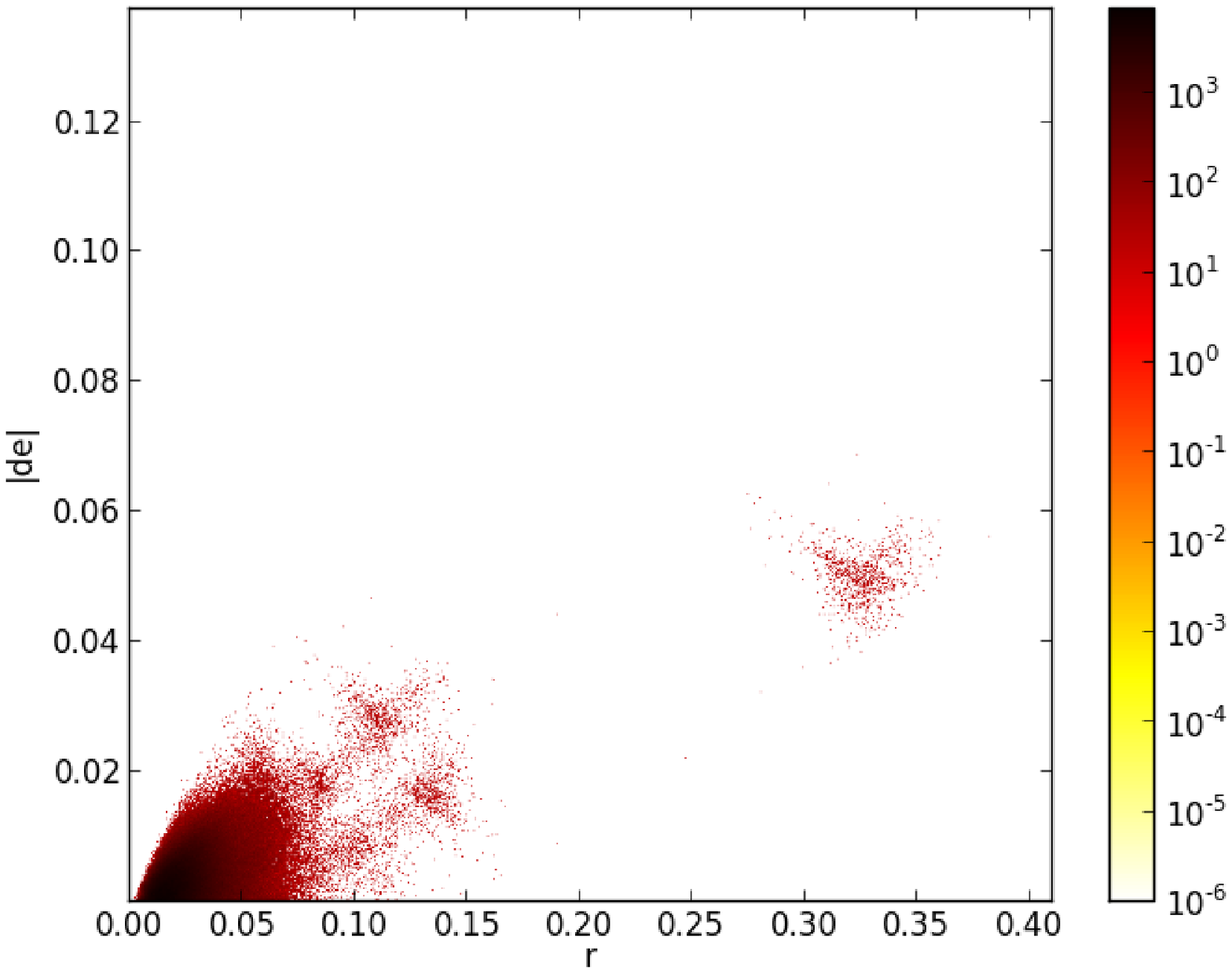}
	\includegraphics[width=0.9\columnwidth]{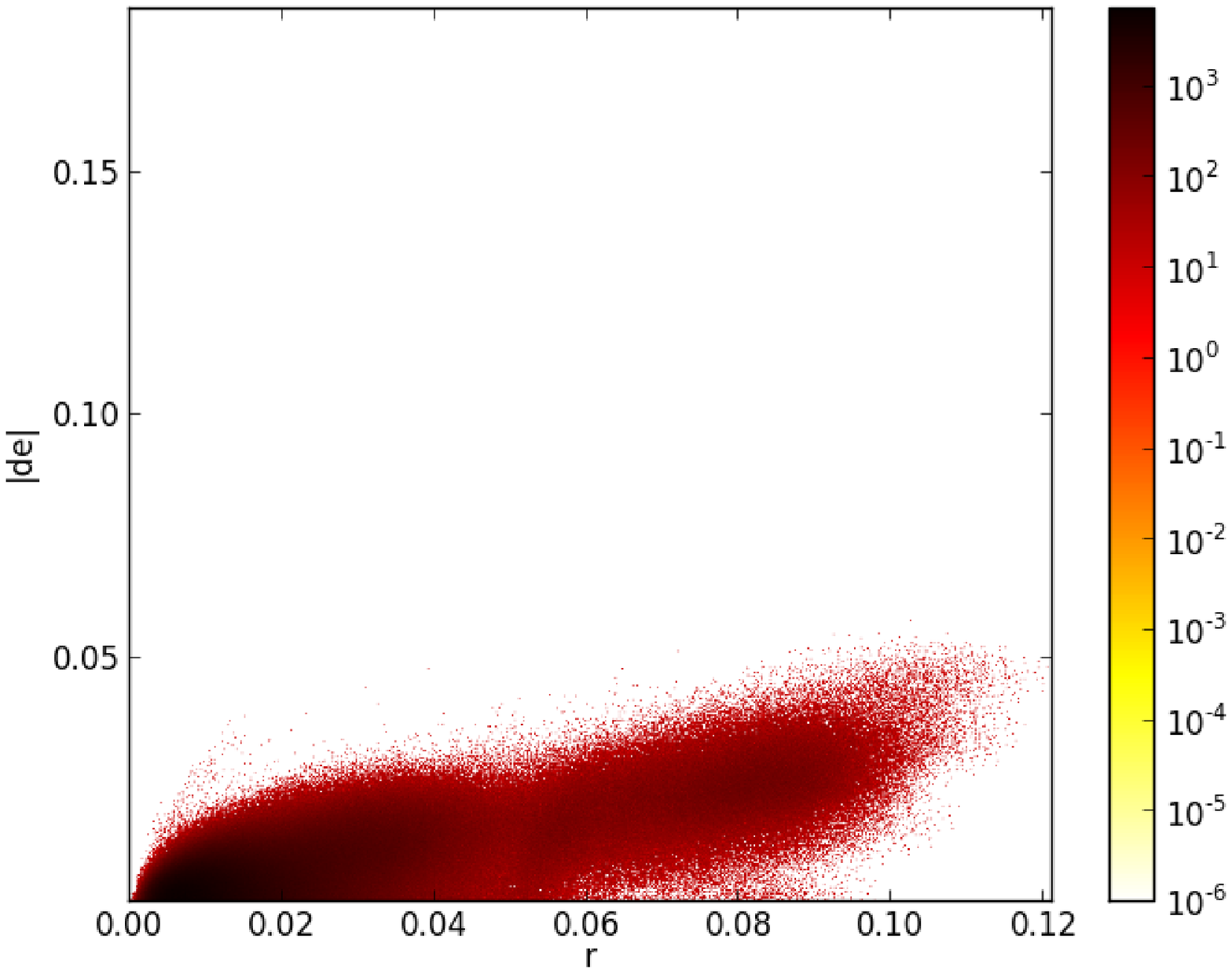}
	\caption{(Color online) The scatter plot energy difference $|d e|$ vs. lattice distance $r$, $d=8$ (top) and $d=13$ (bottom). The plots are generated from random subsets of extreme lattices of $1000$ ($d=8$) an $2000$ ($d=13$) lattices.}
	\label{fig:de-vs-adr-d8-13}
\end{figure}

\begin{figure}
	\includegraphics[width=0.9\columnwidth]{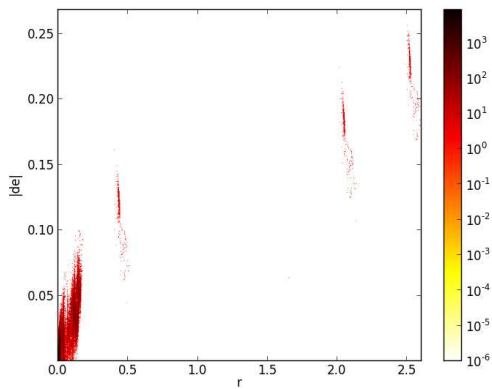}
	\caption{(Color online) The scatter plot energy difference $|d e|$ vs. lattice distance $r$ for $d=17$. The plot is generated from the full set of extreme lattices ($\sim867$) available in that dimension.}
	\label{fig:de-vs-r-d17}
\end{figure}

We have studied correlations between energy difference $|e(\Lambda) - e(\Lambda^\prime)|$ and inter-lattice distance $\rho(\Lambda, \Lambda^\prime)$ for extreme lattices by looking at the scatter plots for $d=8-13$. The $d=8$ and $d=13$ are shown in Fig.~\ref{fig:de-vs-adr-d8-13}. We see that lattices that are \emph{equally dense} are also located very \emph{close} in the space of lattices. We also provide the scatter plot in $d=17$ on Fig.~\ref{fig:de-vs-r-d17} for illustration purposes. We see that the universal trend is present in this case also.

\section{Conclusions and open questions}
\label{sec:conclusions}

In this paper we have studied some statistical properties of extreme lattices generated by the Voronoi algorithm (supplemented with an eutaxy test). In particular we have defined and studied their decorrelation properties (appropriately measured), their symmetries and their distribution of packing fractions and kissing numbers. We have also studied the correlations between these quantities showing how the strong correlation between large packing fraction (kissing number) and symmetries or correlations is diminished in high dimensions.

We have verified that the least dense of the extreme lattices is $A_d$ in all dimensions and for all extreme lattices we have found. We have also seen how for $d\geq 9$ the least dense extreme lattice has considerably more symmetries that the most dense.

We had to stop our analysis at $d=19$, as the generation of extreme lattices becomes impossible with current means. This is probably related to a similar phenomenon found in~\citep{marcotte2013efficient,kallus2013statistical} although the three algorithms are unrelated to each other. We hope to investigate this and the remaining issues in the near future.

Our results apply to relatively low dimensions, $d\leq19$. However, we have discovered that there clearly exist special dimensions, where decorrelation properties are different from the neighboring dimensions. It is of great interest to understand how these results carry on in higher dimensions.

As we have discussed in the introduction, the dimensions we have looked at, $d=8-19$, are very low and our hope was to spot patterns in the properties of extreme lattices, that are valid in higher dimensions as well. 

Our results suggest naturally a number of open problems. We have studied the properties of extreme lattices, like their density and kissing numbers in $d=8-19$. What is the behavior of the typical density and kissing number in higher dimensions and their asymptotic behavior as $d\to\infty$? In Ref.~\onlinecite{andreanov2012random}, we conjectured that typical perfect lattices might improve the Minkowski lower bound on the density. We have found, that the density of typical extreme lattices is bigger than that of typical perfect lattices. Do typical extreme lattices provide (further) improvement of the lower bound? What are the decorrelation properties of $A_d$ and $D_d$ lattices in high dimensions and how do they compare with those of the densest lattices? Do the $A_d$/$D_d$ always decorrelate slower than the typical extreme lattices? Distinct lattices with equal densities have small separation in the space of lattices (see. Figs.~\ref{fig:de-vs-adr-d8-13} and \ref{fig:de-vs-r-d17}), as we have discovered. Is this statement true in all dimensions?

We have seen a special case, $d=13$, where the densest lattice is much more decorrelated, than the densest lattices in $d=12,14$. What is the fate of such exceptional dimensions as $d\to\infty$? That is which of the scenarios proposed in Ref.~\onlinecite{scardicchio2008estimates} (i.e. special dimensions vanish or they persist) is realized for $d\to\infty$?

The decorrelation properties of the lattices were extracted from the pair correlator $g_2$ only. Higher order correlation functions, like $g_3$ or $g_4$, are also important in the context of the decorrelation principle. A more stringent test is to check explicitly whether $g_3$ and $g_4$ factorize into products of $\rho$, the number density of the lattice, and $g_2$. So far such test was only carried out for the "ghost" RSA~\citep{torquato2006exactly}, where all the correlators can be computed exactly.

When studying the symmetries of (extreme) lattice packings, we used the simplest possible measure - the number of the symmetries. There are other measures, which might provide additional information. For example, the $A_d$ and $D_d$ lattices have very large symmetry groups, while the densest known lattices have symmetry groups that are small, compared to $A_d/D_d$, yet the with a richer structure. Their \emph{richness} is reflected in the running time of the algorithm, that computes the groups: while computing the $\autq{A_{20}}$ is a matter of no time even on a desktop machine, the computation of the symmetry group of $\Lambda_{20}$, the densest known lattice in $d=20$, requires days of computing time on the same machine.

Indeed, the size of the automorphism groups of $A_d$ and $D_d$ follow a simple recurrence law (see Appendix B) testifying the simplicity of the groups.

Another interesting problem is how the symmetry of a typical packing is related to its decorrelation properties as $d\to\infty$. Our results for extreme lattices suggest that packings with lower symmetry decorrelate faster. It will be extremely interesting to check this statement with a sphere packing, where we can vary its symmetry at will, and see what is the effect of such variation on decorrelation behavior. Finally, we have only studied the lattice packings. It will be extremely interesting to look at packings with many particles in the unit cell (i.e. periodic packings) and check if our conclusions apply as well to periodic packings or, eventually, how they are altered.

\begin{acknowledgments}
	We would like to thank Achill Sch\"urmann for useful discussions. We acknowledge developers of the libraries PARI~\citep{PARI2} and GNU GSL~\citep{gsl} which were used in simulations. S.T. was supported in part by the National Science Foundation under Grants DMR-$0820341$ and No. DMS-$1211087$. This work was partially supported by a grant from the Simons Foundation (Grant No. $231015$ to Salvatore Torquato).
\end{acknowledgments}
 
\section*{Appendix A. Definitions of metric in the space of lattices}

We used a $g_2$ based metric to compute distances between lattices. Here we discuss possible alternatives and their drawbacks.

The most tempting way to estimate the inter-lattice distance is to use one of the many matrix distances for the lattice Gram matrices. However, this is not a correct definition, since a single lattice can be represented by many different Gram matrices (isometries). It is possible to amend the matrix distance and make it aware of the isometries. We define a distance between two lattices $\Lambda$ and $\Lambda^\prime$ as a minimum matrix distance over all equivalent representations of the two lattices:
\begin{gather}
	\rho(\Lambda,\Lambda^\prime) = \min_{U\in\text{GL}_d(\mathbb{Z})}||U^t\,Q\,U -Q^\prime||^2.	
\end{gather}
where $Q$ and $Q^\prime$ are the respective Gram matrices. The minimization takes into account all possible isometric copies of $\Lambda$.  This is the most straightforward definitions of distance, but it is not practical and we do not know of an implementation of this algorithm. 

Also, any metric that requires the knowledge of a lattice theta series is going to be impractical. There is an algorithm to compute the entire theta series starting from some initial part of the series. The algorithm makes use of theory of modular forms~\citep{conway1999sphere,elkies2000lattices}. The computation is not simple. \textrm{MAGMA} has routines that are able to perform at least part of the computation but the complexity of such computation grows quickly with the number of dimensions, and already in as low as $d=6$ it can be quite involved. 

\section*{Appendix B. Some known automorphism groups of Lattices}

We give below the sizes $|Aut|$ of the groups of automorphisms for the $A_d$ and $D_d$ lattices as well as the densest lattices in dimensions $d=2-14$.

\begin{table}[htbp]
\begin{tabular}{ | c | c | }
\hline
$A_{2}$ & 12 \\
\hline
$A_{3}$ & 48 \\
\hline
$A_{4}$ & 240 \\
\hline
$A_{5}$ & 1440 \\
\hline
$A_{6}$ & 10080 \\
\hline
$A_{7}$ & 80640 \\
\hline
$A_{8}$ & 725760 \\
\hline
$A_{9}$ & 7257600 \\
\hline
$A_{10}$ & 79833600 \\
\hline
$A_{11}$ & 958003200 \\
\hline
$A_{12}$ & 12454041600 \\
\hline
$A_{13}$ & 174356582400 \\
\hline
$A_{14}$ & 2615348736000 \\
\hline
$A_{15}$ & 41845579776000 \\
\hline
$A_{16}$ & 711374856192000 \\
\hline
$A_{17}$ & 12804747411456000 \\
\hline
$A_{18}$ & 243290200817664000 \\
\hline
$A_{19}$ & 4865804016353280000 \\
\hline
$A_{20}$ & 102181884343418880000 \\
\hline
$A_{21}$ & 2248001455555215360000 \\
\hline
$A_{22}$ & 51704033477769953280000 \\
\hline
$A_{23}$ & 1240896803466478878720000 \\
\hline
$A_{24}$ & 31022420086661971968000000 \\
\hline
\end{tabular}
	\caption{The sizes of automoriphism groups for the $A_d$ lattices, $d=2-24$. The number of symmetries is growing factorially in $d$.}
	\label{tab:Ad-aut}
\end{table}

The $D_d$ family of lattice is defined as the set of integer points $\mathbf{x}\in\mathbb{Z}^d$, such that $\sum_i x_i\equiv 0(\mod 2)$~\citep{conway1999sphere}. In $d=3$, the $D_3$ lattice is the FCC lattice. The $D_d$ family of lattices represents the densest lattices in $d=3,4,5$.

\begin{table}[htbp]
\begin{tabular}{ | c | c | }
\hline
$D_{3}$ & 48 \\
\hline
$D_{4}$ & 1152 \\
\hline
$D_{5}$ & 3840 \\
\hline
$D_{6}$ & 46080 \\
\hline
$D_{7}$ & 645120 \\
\hline
$D_{8}$ & 10321920 \\
\hline
$D_{9}$ & 185794560 \\
\hline
$D_{10}$ & 3715891200 \\
\hline
$D_{11}$ & 81749606400 \\
\hline
$D_{12}$ & 1961990553600 \\
\hline
$D_{13}$ & 51011754393600 \\
\hline
$D_{14}$ & 1428329123020800 \\
\hline
$D_{15}$ & 42849873690624000 \\
\hline
$D_{16}$ & 1371195958099968000 \\
\hline
$D_{17}$ & 46620662575398912000 \\
\hline
$D_{18}$ & 1678343852714360832000 \\
\hline
$D_{19}$ & 63777066403145711616000 \\
\hline
$D_{20}$ & 2551082656125828464640000 \\
\hline
$D_{21}$ & 107145471557284795514880000 \\
\hline
$D_{22}$ & 4714400748520531002654720000 \\
\hline
$D_{23}$ & 216862434431944426122117120000 \\
\hline
$D_{24}$ & 10409396852733332453861621760000 \\
\hline
\end{tabular}
	\caption{The sizes of automoriphism groups for the $D_d$ lattices, $d=2-24$. The number of symmetries is growing factorially in $d$.}
	\label{tab:Dd-aut}
\end{table}

The size of their automorphism group is, by inspection of the above tables, 
\begin{eqnarray}
	|Aut(A_d)| & = & 2(d+1)!\ ,\\
	|Aut(D_d)| & = & 2^d d!\ ,
\end{eqnarray}
(the only exception being $D_4$ with the size $1152$ instead of $384$ given by the above formula), giving the following  asymptotic $d\to\infty$ behavior
\begin{eqnarray}
	s(A_d) & \simeq & \ln(d)-1+\Ord{\ln d/d},\\
	s(S_d) & \simeq & \ln(d)+\ln(2)-1+\Ord{\ln d/d}.
\end{eqnarray}
This means that these lattices have a superexponential growth of the size of their automorphism groups, while the best packers have relatively small sized groups, an indication that the decorrelation principle might be at work here.

All the densest lattices represented in Table~\ref{tab:Ld-aut} all belong to the so-called laminated family of lattices, which is is denoted as $\Lambda_d$. These lattices are constructed in a recursive way, starting from $d=1$. Their construction exploits a natural idea, that we can get a dense $d+1$ lattice $\Lambda_{d+1}$ from a dense $d$ one $\Lambda_d$, by stacking the layers of $\Lambda_d$ in a smart way. Namely, we should place the spheres of the next layer in the \emph{deep holes} of the current layer. The deep holes are the points of space that maximize the distance from a point in space to any lattice point. We illustrate this construction, by showing the first few steps of the recursion. In $d=1$ there is a single lattice: it is simply a chain of touching spheres. In $d=2$ we stack the layers, so that every next layer has its spheres shifted by half a lattice spacing with respect the previous layer. This generates the $\Lambda_2\sim A_2$ lattice. Repeating the procedure for the triangular lattice, we get the FCC lattice $A_3\sim D_3$. Notice, that the construction outcome is not unique: it is easy convince yourself that there are uncountably many ways to stack the layer of $A_2$, all of them giving sphere packings of the same density as that of FCC lattice $A_3$. This is an important observation, since in higher dimension the construction procedure also generates many lattices, and one has to pick the densest.

\begin{table}[htbp]
\begin{tabular}{ | c | c | }
\hline
$A_2$ & 12 \\
\hline
$A_3\sim D_3$ & 48 \\
\hline
$D_4$ & 1152 \\
\hline
$D_5$ & 3840 \\
\hline
$E_6\sim\Lambda_{6}$ & 103680 \\
\hline
$E_7\sim\Lambda_{7}$ & 2903040 \\
\hline
$E_8\sim\Lambda_{8}$ & 696729600 \\
\hline
$\Lambda_{9}$ & 10321920 \\
\hline
$\Lambda_{10}$ & 884736 \\
\hline
$K_{11}$ & 207360 \\
\hline
$K_{12}$ & 78382080 \\
\hline
$K_{13}$ & 622080 \\
\hline
$\Lambda_{14}$ & 884736 \\
\hline
$\Lambda_{15}$ & 41287680 \\
\hline
$\Lambda_{16}$ & 89181388800 \\
\hline
$\Lambda_{17}$ & 1486356480 \\
\hline
$\Lambda_{18}$ & 159252480 \\
\hline
$\Lambda_{19}$ & 23592960 \\
\hline
$\Lambda_{24}$ & 8315553613086720000 \\
\hline
$Q_{32}$ & 207360 \\
\hline
\end{tabular}
	\caption{The sizes of automoriphism groups for the densest known lattices in $d=2-19$ and $d=24$. Unlike the cases of $A_d$ and $D_d$, the growth is not monotonic but rather it shows irregular oscillations on an underlying growing trend.}
	\label{tab:Ld-aut}
\end{table}

\bibliography{packing}{}

\end{document}